\documentclass[10pt,journal,comsoc]{IEEEtran}
\usepackage{graphicx} 
\usepackage{graphics}
\usepackage{multirow}
\usepackage{textcomp}
\usepackage{array}
\usepackage{xcolor}

\usepackage{tabularx}
\newcolumntype{Y}{>{\centering\arraybackslash}X} 
\usepackage{booktabs}

\usepackage{caption}
\usepackage{subcaption}
\usepackage{multirow}

\makeatletter
\renewcommand\p@subfigure{\thefigure} 
\makeatother

\usepackage{tikz}
\usetikzlibrary{shapes.geometric, arrows.meta, positioning, fit, calc, backgrounds, shadows}

\usepackage{amsmath,amssymb,amsfonts}
\usepackage{bm}

\usepackage[algo2e,linesnumbered,ruled]{algorithm2e} 

\usepackage{amsthm}
\theoremstyle{definition}

\newtheorem{definition}{Definition}

\newtheorem{proposition}{Proposition}

\newtheorem{remark}{Remark}

\SetKwProg{Fn}{Function}{}{end}

\SetKwInput{KwInit}{Init}
\SetAlgoSkip{}

\newcolumntype{M}[1]{>{\centering\arraybackslash}m{#1}}

\DeclareSymbolFont{symbolsC}{U}{txsyc}{m}{n}
\DeclareMathSymbol{\notniFromTxfonts}{\mathrel}{symbolsC}{61}

\usepackage{titlesec}

\titlespacing{\section}{0pt}{6pt plus .2ex minus .2ex}{0.5ex plus .2ex}
\titlespacing{\subsection}{0pt}{6pt plus .2ex minus .2ex}{0.5ex plus .2ex}

\title{Spatiotemporal Continual Federated Learning for Agentic Multi-UAV Edge Networks: Mitigating Catastrophic Forgetting}

\author{Chuan-Chi~Lai,~\IEEEmembership{Member,~IEEE}
    \IEEEcompsocitemizethanks{
        \IEEEcompsocthanksitem{
        This work was supported by the National Science and Technology Council (NSTC), Taiwan, under Grant No. NSTC 115-2221-E-194-042-MY2, and was also supported in part by the Advanced Institute of Manufacturing with High-tech Innovations (AIM-HI) from the Featured Areas Research Center Program within the framework of the Higher Education Sprout Project by the Ministry of Education (MOE) in Taiwan. \emph{(Corresponding author: Chuan-Chi~Lai.)}}
        \IEEEcompsocthanksitem{Chuan-Chi Lai is with the Department of Communications Engineering, National Chung Cheng University, Minxiong Township, Chiayi County 621301, Taiwan, and also with the Advanced Institute of Manufacturing with High-tech Innovations (AIM-HI), National Chung Cheng University, Minxiong Township, Chiayi County 621301, Taiwan (e-mail: chuanclai@ccu.edu.tw).}
    }
}

\IEEEtitleabstractindextext{%
\begin{abstract}
This paper addresses multi-objective conflicts and catastrophic forgetting in uncrewed aerial vehicle (UAV) networks across dynamic spatiotemporal environments. Conventional multi-agent reinforcement learning (MARL) algorithms suffer from severe policy degradation during sequential task transitions. We propose a spatiotemporal continual federated learning (SCFL) framework driven by the group-decoupled multi-agent proximal policy optimization (G-MAPPO) algorithm. SCFL incorporates a three-stage geometric alignment mechanism: it resolves local gradient conflicts via group-decoupled policy optimization (GDPO), mitigates spatial Non-Independent and Identically Distributed (non-IID) client drift through adaptive cosine aggregation, and suppresses inter-task interference via global temporal orthogonal projection without raw experience replay. Evaluations show that SCFL achieves superior robustness over federated baselines, maintaining spatial service reliability above 0.95 and a load balancing index of approximately 0.95 during non-stationary transitions. A longitudinal self-degradation analysis further shows that SCFL preserves historical knowledge with near-zero performance variation in spatial reliability and QoS under moderate loads from 40 to 120 users, while revealing its operating boundary under extreme congestion with 140 users due to hard projection constraints. The framework provides a scalable, communication-efficient approach for autonomous aerial network orchestration.
\end{abstract}

}

\begin{document}

\setlength{\abovedisplayskip}{3pt}
\setlength{\belowdisplayskip}{3pt}
\setlength{\abovedisplayshortskip}{3pt}
\setlength{\belowdisplayshortskip}{3pt}

\setlength{\textfloatsep}{6pt plus 2pt minus 2pt}
\setlength{\floatsep}{6pt plus 2pt minus 2pt}
\setlength{\intextsep}{6pt plus 2pt minus 2pt}

\maketitle
\IEEEdisplaynontitleabstractindextext

%
\IEEEpeerreviewmaketitle

\IEEEPARstart{I}{n} recent years, deploying \textit{Uncrewed Aerial Vehicles} (UAVs) as \textit{Aerial Base Stations} (ABSs) has emerged as a practical solution for enhancing the coverage and capacity of next-generation wireless networks~\cite{8660516}. Compared to conventional terrestrial infrastructure, ABSs offer high mobility and the ability to establish \textit{Line-of-Sight} (LoS) communication links through flexible 3D positioning~\cite{al2014modeling}. These characteristics make them suitable for addressing temporary traffic surges~\cite{8642333}, restoring emergency communications, and bridging coverage gaps in remote regions~\cite{11119219,11051254,9453853}. Despite their potential, orchestrating agentic multi-UAV swarms (where individual nodes autonomously perceive, learn, and adapt to environments without centralized control) faces significant challenges stemming from dynamic and non-stationary user distributions. Mobile traffic exhibits strong spatiotemporal tidal effects, shifting drastically between dense urban business districts and sparse suburban residential areas~\cite{7917576}.

Traditional \textit{Multi-Agent Reinforcement Learning} (MARL) algorithms navigating these transitions frequently suffer from catastrophic forgetting. Agents overwrite previously learned optimal policies when adapting to new environments, which can degrade performance and service stability~\cite{pnas2017}. Furthermore, edge artificial intelligence systems require continual learning to maintain operational effectiveness in dynamic environments~\cite{11363404}. For instance, an agentic UAV swarm in a dense urban area must adjust transmission power and 3D positioning to mitigate co-channel interference. Conversely, in a sparse rural scenario, agents adjust strategies to expand coverage. Catastrophic forgetting manifests when the swarm, upon adapting to rural environments, loses its urban interference management policies. Returning to an urban cluster, the application of aggressive rural policies may cause interference issues and network congestion.

This challenge is compounded by the need to simultaneously balance heterogeneous objectives, including energy efficiency, user fairness, and minimum \textit{Quality of Service} (QoS) guarantees. These objectives possess conflicting gradients and varying physical magnitudes, which can destabilize the learning process through destructive interference~\cite{6918520}. Recent decentralized approaches mitigate forgetting by optimizing network connectivity, such as introducing logical teleportation links to accelerate knowledge sharing~\cite{11168215}. However, existing strategies typically rely on periodic offline retraining or transfer learning~\cite{10657767}. These methods incur computational overhead and latency, making them less suitable for real-time online coordination.

To bridge this gap, this paper proposes the \textit{Spatiotemporal Continual Federated Learning} (SCFL) framework, which integrates a decentralized federated architecture with a \textit{Group-Decoupled Policy Optimization} (GDPO)~\cite{liu2026gdpogrouprewarddecouplednormalization} enhanced \textit{Multi-Agent Proximal Policy Optimization} (MAPPO)~\cite{NIPS2022_1787_Yu} algorithm, termed G-MAPPO. Rather than requiring computationally expensive offline retraining, our framework enables the distributed UAV swarm to adapt continuously to spatiotemporal variations while preserving data privacy. The primary contributions of this work are summarized as follows:
\begin{itemize}
    \item \textbf{Spatiotemporal Decoupled Alignment Architecture}: We propose a three-stage geometric alignment framework that separates multi-objective conflict resolution, spatial \textit{Non-Independent and Identically Distributed} (non-IID) aggregation, and temporal knowledge retention. Specifically, GDPO is employed to normalize heterogeneous reward scales locally and resolve gradient conflicts, while spatial cosine aggregation mitigates client drift across edge nodes.
    \item \textbf{Temporal Knowledge Retention via Gradient Projection}: To mitigate catastrophic forgetting during sequential task transitions (Urban, Suburban, and Rural), we enforce global temporal orthogonal gradient projection using compressed historical anchor gradients. We theoretically establish first-order non-interference with stored historical anchor gradients, avoiding the backhaul and storage overhead associated with raw experience replay.    
    \item \textbf{Empirical Verification and Boundary Analysis}: We systematically evaluate the framework across dynamic task transitions and varying user densities ranging from 40 to 140 users. Horizontally, SCFL mitigates reward collapse and maintains an average load balancing index of $\approx 0.95$. Longitudinally, a self-degradation analysis demonstrates near-zero performance variation in spatial reliability and QoS under moderate loads, while revealing the empirical operating boundary of hard projection constraints under extreme congestion with 140 users.
\end{itemize}

The remaining sections of this paper are organized as follows: Section~\ref{sec:related_work} reviews related work on UAV deployment and continual learning. Section~\ref{sec:system_model} presents the system model and problem formulation. Section~\ref{sec:proposed_framework} details the SCFL framework. Section~\ref{sec:theoretical_analysis} provides the theoretical analysis of knowledge retention. Section~\ref{sec:simulation_results} discusses the simulation setup and performance evaluation. Finally, Section~\ref{sec:conclusion} concludes the paper.

\section{Related Work}
\label{sec:related_work}

\subsection{UAV Deployment and 3D Trajectory Design}

UAV deployment optimization has been extensively investigated to maximize coverage probability and spectral efficiency in wireless networks. Early analytical channel modeling studies established fundamental relationships between UAV altitude and air-to-ground path loss probabilities~\cite{al2014modeling}. Subsequent research focused on optimizing 3D placement configurations to decouple coverage and capacity constraints in heterogeneous cellular architectures~\cite{7486987}, addressing coverage overlapping for arbitrary user crowds~\cite{9177297}, and designing energy-efficient trajectories that balance propulsion power consumption with throughput demands~\cite{7888557,9946428}. 

Recent studies further extend these concepts to complex 3D scenarios by formulating mixed-integer nonconvex optimization problems to minimize energy consumption while determining the optimal number of deployed UAVs~\cite{Gong2024Energy}. To address operational hazards, interference-aware path planning strategies have been developed to mitigate line-of-sight signal interference~\cite{8654727}. Furthermore, advanced evolutionary algorithms have been proposed for dynamic multi-objective path planning in urban environments, balancing energy consumption, collision risks, and path length~\cite{Xu2025Evolving}. 

However, conventional optimization approaches typically rely on convex relaxation or heuristics that require perfect \textit{Channel State Information} (CSI) and assume static user distributions. Although recent frameworks incorporate 3D mobility, altitude is frequently treated either as a fixed parameter or as an independent optimization variable decoupled from horizontal adjustments~\cite{7918510}. Consequently, these models lack the tight spatiotemporal coupling required for autonomous network adaptation under rapid, non-stationary user density variations~\cite{7917576}.

\subsection{Multi-Agent Reinforcement Learning for UAV Swarms}

To manage the high environmental complexity and operational dynamics of UAV swarms, \textit{Multi-Agent Reinforcement Learning} (MARL) has been extensively adopted. Decentralized cooperative policies for interference management and trajectory control are frequently formulated using algorithms such as \textit{Multi-Agent Deep Deterministic Policy Gradient} (MADDPG)~\cite{lowe2017multi}, as demonstrated in~\cite{10373024,9779087}. More recently, MAPPO~\cite{NIPS2022_1787_Yu} has gained prominence due to its on-policy update stability in cooperative tasks. To address scalability during dynamic cluster reconfiguration, hierarchical multi-agent reinforcement learning frameworks have been introduced to jointly optimize power allocation and mobility management~\cite{Meer2025Hierarchical}. 

In addition, spatiotemporal-aware architectures incorporating transformer mechanisms have been developed to capture environmental dynamics and guarantee deterministic communication requirements during cooperative coverage operations~\cite{Chen2026Spatiotemporal,8807386}. Despite these advancements, standard MARL formulations face fundamental challenges when simultaneously optimizing heterogeneous metrics such as fairness, delay, and energy efficiency. Conflicting gradients among these objectives often introduce severe training instability~\cite{6918520}. While hybrid approaches combining Lyapunov optimization with reinforcement learning address queue stability under varying traffic demands~\cite{Hoang2025Adaptive}, they typically rely on stationary model-based constraints and suffer from severe policy degradation when user distributions shift dynamically.

\begin{table*}[t]
\caption{Qualitative Comparison of the Proposed SCFL Framework with Existing Literature}
\label{tab:qualitative_comparison}
\centering
\begin{tabular}{@{}lccccc@{}}
\toprule
\textbf{Reference} & \textbf{3D Trajectory} & \textbf{Multi-Agent} & \textbf{Multi-Objective} & \textbf{Spatial Non-IID} & \textbf{Continual} \\
 & \textbf{Optimization} & \textbf{Collaboration} & \textbf{Conflict Resolution} & \textbf{Alignment} & \textbf{Learning} \\ \midrule
Mozaffari \textit{et al.}~\cite{7486987} & $\times$ & $\times$ & $\times$ & $\times$ & $\times$ \\
Zeng \textit{et al.}~\cite{7888557} & \checkmark & $\times$ & $\times$ & $\times$ & $\times$ \\
Liu \textit{et al.}~\cite{10373024} & \checkmark & \checkmark & $\times$ & $\times$ & $\times$ \\
Chen \textit{et al.}~\cite{Chen2025ConMod} & \checkmark & \checkmark & $\times$ & $\times$ & \checkmark \\
Chen \textit{et al.}~\cite{Chen2026HierFCL} & $\times$ & \checkmark & $\times$ & \checkmark & \checkmark \\ \midrule
\textbf{Proposed SCFL Framework} & \checkmark & \checkmark & \checkmark & \checkmark & \checkmark \\ \bottomrule
\end{tabular}
\vspace{-1em}
\end{table*}

\subsection{Continual Learning and Adaptation in Wireless Networks}

\textit{Continual Learning} (CL) methodologies specifically target the stability-plasticity dilemma in non-stationary systems~\cite{sp_dilemma_1988}. When migrating continual learning into decentralized federated edge networks (i.e., federated continual learning), current techniques for mitigating catastrophic forgetting face unique hurdles. Regularization approaches, such as \textit{Elastic Weight Consolidation} (EWC)~\cite{pnas2017}, impose rigid penalties on weight updates to preserve prior knowledge; however, such inflexible constraints may limit the parameter plasticity required for rapid adaptation under large distribution shifts in highly dynamic aerial environments. Alternatively, replay-based methods rely on storing and periodically retraining on historical raw datasets. In multi-UAV federated edge networks, transmitting and storing high-dimensional raw trajectories or experience replay buffers across resource-constrained wireless links raises substantial backhaul, storage, and privacy concerns. Furthermore, while centralized gradient projection baselines like \textit{Gradient Episodic Memory} (GEM)~\cite{GEM_citation} protect past knowledge manifolds, they require centralized access to raw historical samples, creating substantial communication, storage, and privacy challenges in federated aerial networks.

Recent wireless literature has explored CL in various dynamic environments, including terrestrial resource allocation via graph neural networks~\cite{Zhou2025CLGNN}, satellite channel prediction~\cite{Wang2026CLLEO}, and cross-domain edge intelligence~\cite{Hu2026EdgeCL}. In multi-agent scenarios, specialized CL frameworks have been developed to handle dynamic 3D aerial perspectives for object recognition~\cite{Chen2025ConMod} and to align spatial non-IID distributions via hierarchical federated clustering in vehicular networks~\cite{Chen2026HierFCL}. 

Nevertheless, applying continual learning to multi-objective resource allocation in 3D UAV networks remains largely unexplored. Existing solutions predominantly optimize single objectives or focus on isolated tasks such as image recognition, lacking mechanisms to resolve gradient conflicts among diverse communication metrics or to couple 3D spatial non-IID client drift with temporal catastrophic forgetting. 

To bridge this critical gap, the proposed SCFL framework adopts a gradient projection paradigm tailored specifically for federated aerial edge networks. Unlike conventional replay techniques requiring massive storage or centralized sample access, our approach extracts a compressed reference anchor gradient at the global server at the end of each task phase. By enforcing temporal orthogonal gradient projection at the global parameter server alongside local multi-objective and spatial non-IID alignments, SCFL prevents destructive interference with historical task manifolds without incurring the communication overhead or privacy violations associated with raw experience replay. 

Table~\ref{tab:qualitative_comparison} highlights the contributions of this study by qualitatively comparing the proposed SCFL framework with representative literature across key technical dimensions.

\section{System Model and Problem Formulation}
\label{sec:system_model}

To model decentralized UAV swarms, we adopt a federated edge architecture where distributed UAVs act as clients communicating with a central server. Our focus is not on generic federated aggregation protocol design; rather, we tailor the aggregation mechanism to mitigate spatial non-IID drift within the proposed continual learning framework.

\subsection{3D Aerial-Ground Network Architecture}

\begin{figure}[!t] 
    \centering
    \includegraphics[width=\columnwidth]{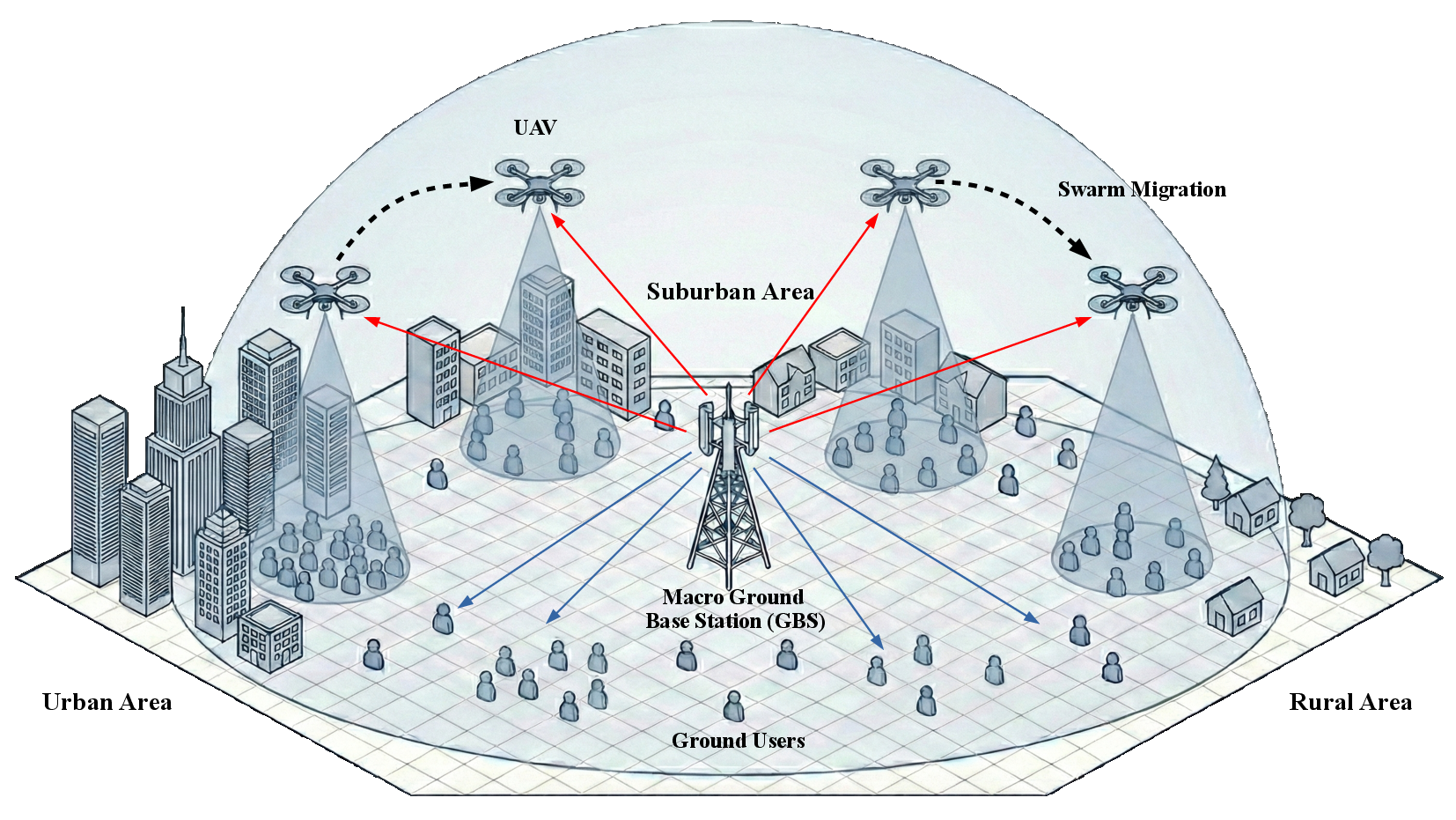} 
    \caption{Illustration of the 3D aerial-ground integrated network deployed across a heterogeneous environment. The service area features a spatial transition from a dense urban area (left), through a suburban area (center), to a sparse rural area (right). A central macro GBS provides ubiquitous omnidirectional coverage (hemispherical dome), while multiple UAVs function as mobile small cells. The dashed arrows illustrate the spatiotemporal migration of the UAV swarm, emphasizing its continuous adaptation to shifting user densities across the sequential task chain.}
    \label{fig:system_model}
\end{figure}

As illustrated in Fig.~\ref{fig:system_model}, we consider a downlink aerial-ground integrated network comprising $N$ rotary-wing UAVs functioning as aerial base stations and $M$ ground users distributed over a geographical area $\mathcal{D} \subset \mathbb{R}^2$. The set of UAVs is denoted by $\mathcal{U} = \{1, \dots, N\}$, and the set of ground users is denoted by $\mathcal{M} = \{1, \dots, M\}$.

Unlike conventional 2D deployment models, the UAVs possess 3D mobility to facilitate spatial adaptation. The instantaneous position of the $i$-th UAV at time step $t$ is represented by a 3D coordinate vector $\mathbf{q}_i(t) = [x_i(t), y_i(t), h_i(t)]^T$, where $[x_i(t), y_i(t)]$ denotes the horizontal position and $h_i(t)$ represents the altitude. To ensure operational safety, the altitude is constrained within a predefined range $H_{\min} \leq h_i(t) \leq H_{\max}$. This vertical degree of freedom allows the network to dynamically adjust the service footprint in response to varying user densities across the heterogeneous environment.

To provide ubiquitous coverage and backhaul support, a macro \textit{Ground Base Station} (GBS) is deployed at the center of the service area at $\mathbf{q}_{\mathrm{GBS}} = [x_{\mathrm{GBS}}, y_{\mathrm{GBS}}, H_{\mathrm{GBS}}]^T$. The GBS operates with a transmit power $P_{\mathrm{GBS}}$ that is significantly higher than the UAV transmit power $P_{\mathrm{UAV}}$, serving as an anchor node for users outside the effective coverage of the UAVs. 

\subsection{Heterogeneous Communication and User Association Model}

For the communication link between the macro GBS and ground user $u$, we adopt a standard terrestrial path loss model that accounts for urban shadowing effects. The path loss $L_{\mathrm{GBS},u}(t)$ in dB is modeled as:
\begin{equation}
    L_{\mathrm{GBS},u}(t) = \mathrm{PL}(d_{\mathrm{ref}}) + 10 \kappa \log_{10}\left(\frac{d_{\mathrm{GBS},u}(t)}{d_{\mathrm{ref}}}\right) + \chi_{\mathrm{sh}},
\end{equation}
where $d_{\mathrm{GBS},u}(t)$ is the Euclidean distance between the GBS and user $u$, $d_{\mathrm{ref}}$ is the reference distance, and $\kappa$ is the path loss exponent. The shadowing term $\chi_{\mathrm{sh}} \sim \mathcal{N}(0, \sigma_{\mathrm{sh}}^2)$ is modeled as a zero-mean Gaussian random variable with standard deviation $\sigma_{\mathrm{sh}}$.

Unlike the UAV links that may  benefit from high LoS probabilities, the GBS link is dominantly NLoS due to low antenna height and dense building blockage. This distinct propagation characteristic motivates the deployment of UAVs to provide coverage extension and capacity offloading for edge users.

The communication links between UAVs and ground users are modeled using a probabilistic \textit{Line-of-Sight} (LoS) channel model. The probability of establishing an LoS link between the $i$-th UAV and the $u$-th user depends on the elevation angle $\psi_{i,u}(t) = \arctan\left(\frac{h_i(t)}{r_{i,u}(t)}\right)$, where $r_{i,u}(t)$ is the horizontal distance. The LoS probability is given by~\cite{al2014modeling}:
\begin{equation}
    P_{\mathrm{LoS}}(\psi_{i,u}(t)) = \frac{1}{1 + a \cdot \exp\left(-b (\psi_{i,u}(t) - a)\right)},
\end{equation}
where $a$ and $b$ are environment-dependent constants. The corresponding \textit{Non-Line-of-Sight} (NLoS) probability is defined as $P_{\mathrm{NLoS}} = 1 - P_{\mathrm{LoS}}$. The average path loss is then formulated as:
\begin{equation}
    \bar{L}_{i,u}(t) = P_{\mathrm{LoS}} \cdot L^{\mathrm{LoS}}_{i,u}(t) + P_{\mathrm{NLoS}} \cdot L^{\mathrm{NLoS}}_{i,u}(t),
\end{equation}
where $L^{\mathrm{LoS}}_{i,u}$ and $L^{\mathrm{NLoS}}_{i,u}$ incorporate the free-space path loss along with additional attenuation factors $\eta_{\mathrm{LoS}}$ and $\eta_{\mathrm{NLoS}}$, respectively.


Each ground user $u$ associates with the node providing the strongest reference signal power, which can be either a UAV or the GBS. Let $k \in \mathcal{U} \cup \{\mathrm{GBS}\}$ denote the serving node. The received \textit{Signal-to-Interference-plus-Noise Ratio} (SINR) for user $u$ at time $t$ is expressed as:
\begin{equation}
    \Gamma_u(t) = \frac{P_k G_{k,u}(t)}{\sigma^2 + \sum_{j \neq k} P_j G_{j,u}(t)},
\end{equation}
where $P_k$ is the transmit power and $\sigma^2$ is the noise power. The term $G_{k,u}(t)$ represents the effective channel gain, defined as:
\begin{equation}
    G_{k,u}(t) = G_k^{\mathrm{ant}} \cdot 10^{-\bar{L}_{k,u}(t)/10},
\end{equation}
where $\bar{L}_{k,u}(t)$ is the path loss derived in the previous subsections. Specifically, we assume the GBS employs a static omnidirectional antenna with a constant gain $G_{\mathrm{GBS}}^{\mathrm{ant}}$, while UAVs are equipped with downlink antennas having gain $G_{\mathrm{UAV}}^{\mathrm{ant}}$. 

The achievable downlink data rate for user $u$ at time $t$ is formulated according to the Shannon capacity bound as $R_u(t) = B \log_2(1 + \Gamma_u(t)),$ where $B$ represents the system bandwidth allocated to the user.

\subsection{Spatiotemporal User Distribution Models}

To emulate the non-stationary nature of real-world traffic, the macro-level spatial distribution of ground users, denoted by the probability density function $\Phi(\mathbf{p})$, varies according to a sequential task chain. Three distinct spatial models are defined to represent the urban, suburban, and rural environments.

\subsubsection{Crowded Urban Scenario}
The urban environment is characterized by high user density concentrated in specific hotspots. This distribution is modeled using a \textit{Thomas Cluster Process} (TCP)~\cite{haenggi2012stochastic}. In this model, parent points representing cluster centers are generated with intensity $\lambda_p$, and daughter points representing users are distributed around each parent according to an isotropic Gaussian distribution with variance $\sigma_{\mathrm{pop}}^2$. The density function for a user location $\mathbf{p}$ is given by:
\begin{equation}
    \Phi_{\mathrm{U}}(\mathbf{p}) = \frac{1}{C} \sum_{c=1}^{C} \frac{1}{2\pi\sigma_{\mathrm{pop}}^2} \exp \left( - \frac{\| \mathbf{p} - \mathbf{p}_c \|^2}{2\sigma_{\mathrm{pop}}^2} \right),
\end{equation}
where $C$ is the number of hotspots, $\mathbf{p}_c$ denotes the center of the $c$-th cluster, and $\sigma_{\mathrm{pop}}$ controls the spread of the cluster.

\subsubsection{Suburban Scenario}
The suburban environment represents a transition state with moderate user density. This phase features a combination of residential clusters and scattered users, modeled using a \textit{Gaussian Mixture Model} (GMM) combined with a uniform background component:
\begin{equation}
    \Phi_{\mathrm{S}}(\mathbf{p}) = \alpha \cdot \frac{1}{|\mathcal{D}|} + (1-\alpha) \sum_{c=1}^{C'} \pi_c \mathcal{N}(\mathbf{p} | \boldsymbol{\mu}_c, \boldsymbol{\Sigma}_c),
\end{equation}
where $\alpha \in [0, 1]$ represents the proportion of background users, $|\mathcal{D}|$ is the area of the region, $\pi_c$ is the weight of the $c$-th cluster, and $\mathcal{N}(\cdot)$ denotes the Gaussian density function.

\subsubsection{Rural Scenario}
The rural environment is characterized by sparse user density and a lack of distinct hotspots. The user locations are modeled using a \textit{Homogeneous Poisson Point Process} (HPPP), which results in a uniform distribution over the service area $\mathcal{D}$:
\begin{equation}
    \Phi_{\mathrm{R}}(\mathbf{p}) = \begin{cases}
        \frac{1}{|\mathcal{D}|}, & \text{if } \mathbf{p} \in \mathcal{D}, \\
        0, & \text{otherwise}.
    \end{cases}
\end{equation}

\subsection{Spatiotemporal User Mobility Dynamics}

While the aforementioned distribution models dictate the macroscopic topological states across different phases, the microscopic user mobility during each phase introduces continuous non-stationary dynamics. We adopt a hybrid \textit{Reference Point Group Mobility} (RPGM) and \textit{Gauss-Markov} (GM) mobility model to capture this temporal evolution~\cite{camp2002survey}.

Let $\mathbf{p}_c(t)$ denote the logical reference point of the $c$-th user cluster defined in the TCP or GMM models. These reference points slowly migrate across the service area to simulate shifting hotspots. For each individual user $u$ belonging to cluster $c$, the microscopic velocity $\mathbf{v}_u(t)$ evolves according to the GM process:
\begin{equation}
    \mathbf{v}_u(t) = \alpha_m \mathbf{v}_u(t-1) + (1-\alpha_m)\bar{\mathbf{v}}_c(t) + \sqrt{1-\alpha_m^2} \mathbf{z}_t,
\end{equation}
where $\alpha_m \in [0, 1]$ is the memory parameter governing velocity inertia, $\bar{\mathbf{v}}_c(t)$ is the asymptotic mean velocity driven by the corresponding group reference point $\mathbf{p}_c(t)$, and $\mathbf{z}_t \sim \mathcal{N}(\mathbf{0}, \sigma_z^2 \mathbf{I})$ represents an uncorrelated Gaussian driving process. The position of user $u$ is subsequently updated by $\mathbf{q}_u(t) = \mathbf{q}_u(t-1) + \mathbf{v}_u(t) \Delta t$. This hybrid model ensures that user trajectories exhibit realistic spatial clustering and temporal smoothness, presenting a rigorous continuous adaptation challenge for the UAV swarm.

\subsection{UAV Propulsion Energy Model}

In addition to communication transmit power, the physical flight energy required to sustain 3D mobility is a dominant factor in UAV operations. Based on the aerodynamic principles of rotary-wing aircraft, the propulsion power $P_{\mathrm{fly}}(V_i(t))$ required for the $i$-th UAV flying at a horizontal speed $V_i(t) = \|\mathbf{v}_i(t)\|$ is modeled as~\cite{7888557}:
\begin{align}
    P_{\mathrm{fly}}(V_i(t)) &= P_0 \left(1 + \frac{3 V_i(t)^2}{U_{\mathrm{tip}}^2} \right) + \frac{1}{2} d_{\mathrm{fus}} \rho s_{\mathrm{rotor}} A V_i(t)^3 \notag \\
    &\quad + P_{\mathrm{ind}} \left( \sqrt{1 + \frac{V_i(t)^4}{4 v_0^4}} - \frac{V_i(t)^2}{2 v_0^2} \right)^{1/2},
\end{align}
where $P_0$ and $P_{\mathrm{ind}}$ are the blade profile power and induced power in hovering status, respectively. $U_{\mathrm{tip}}$ denotes the tip speed of the rotor blade, $v_0$ is the mean rotor induced velocity in hover, $d_{\mathrm{fus}}$ is the fuselage drag ratio, $\rho$ is the air density, $s_{\mathrm{rotor}}$ is the rotor solidity, and $A$ is the rotor disc area. The total energy consumption of the network integrates both the dynamic propulsion power and the static transmit power over the operational timeline.

\subsection{Problem Formulation}

The primary objective is to develop a decentralized control policy $\boldsymbol{\pi}$ that addresses the stability-plasticity dilemma in non-stationary environments. The agent must maximize the long-term system utility across a continuous spatiotemporal task chain $\mathcal{T}$, while ensuring that new spatial adaptations do not overwrite consolidated historical knowledge. The optimization problem is mathematically formulated as:
\begin{equation}
\max_{\boldsymbol{\pi}} \mathbb{E} \left[ \sum_{t=0}^{T_{\mathrm{hor}}} \gamma^t U_{\mathrm{total}}(t) \right]
\end{equation}
subject to the following physical and operational constraints:
\begin{align}
    \text{C1: } & H_{\min} \leq h_i(t) \leq H_{\max}, \quad \forall i \in \mathcal{U}, \forall t, \\
    \text{C2: } & \mathbf{q}_i^{xy}(t) \in \mathcal{D}, \quad \forall i \in \mathcal{U}, \forall t, \\
    \text{C3: } & \| \mathbf{q}_i(t) - \mathbf{q}_j(t) \| \geq d_{\min}, \quad \forall i \neq j, \forall t,
\end{align}
where $\gamma \in [0, 1)$ denotes the discount factor.

These constraints are defined as follows: (C1) enforces flight altitude limits to comply with regulatory safety rules; (C2) restricts the horizontal movement of the UAVs to remain within the designated service region $\mathcal{D}$; and (C3) imposes a collision avoidance limit to ensure the minimum safety distance $d_{\min}$ between any two UAVs.

The utility function $U_{\mathrm{total}}$ incorporates conflicting objectives including propulsion-aware energy efficiency, user fairness, and spatial coverage reliability. These objectives are first mapped to the localized RL reward space in Section~\ref{subsec:reward_composition}, while their explicit macroscopic mathematical definitions for system evaluation are detailed in Section~\ref{subsec:setup}. The presence of these diverse metrics necessitates a sophisticated optimization strategy capable of balancing these trade-offs while strictly adhering to safety constraints.

\subsection{Complexity and Methodology Motivation}

The formulated optimization problem is inherently challenging due to its high-dimensional and non-convex objective landscape. The joint optimization of 3D UAV positioning and user association is classified as \textit{Non-deterministic Polynomial-time hard} (NP-hard): the search space for spatial coordinates is continuous in 3D, while the user association represents a large-scale combinatorial sub-problem. 

Traditional optimization techniques typically assume a stationary user distribution and require perfect \textit{Channel State Information} (CSI) to guarantee convergence. However, under the proposed hybrid mobility model, the environment exhibits continuous tidal effects and microscopic velocity inertia. Re-solving the global optimization problem from scratch for every environmental shift leads to prohibitive computational latency, massive backhaul signaling overhead, and a complete loss of temporal experience.

By adopting a federated continual learning approach, specifically the proposed spatiotemporal continual learning framework, the swarm learns a generalized control policy that maps local observations to optimal 3D movements. Compared to traditional optimization and centralized training, this framework provides three core advantages:
\begin{itemize}
    \item \textbf{Online adaptation}: Agents autonomously maneuver to compensate for density fluctuations without requiring a central optimizer to constantly gather global network states.
    \item \textbf{Low-latency and low-overhead execution}: Decentralized execution and federated model aggregation circumvent the continuous transmission of raw privacy-sensitive local trajectories, satisfying the strict timing and backhaul constraints of aerial swarms.
    \item \textbf{Mechanism for knowledge retention}: Through group-decoupled normalization and orthogonal gradient projection, the system mathematically ensures that critical interference management and propulsion optimization skills are not forgotten during long-term non-stationary transitions.
\end{itemize}

\section{Proposed Spatiotemporal Continual Federated Learning Framework}
\label{sec:proposed_framework}

\subsection{Federated CTDE Architecture Overview}

\begin{figure*}[t]
    \centering
    \includegraphics[width=.825\textwidth]{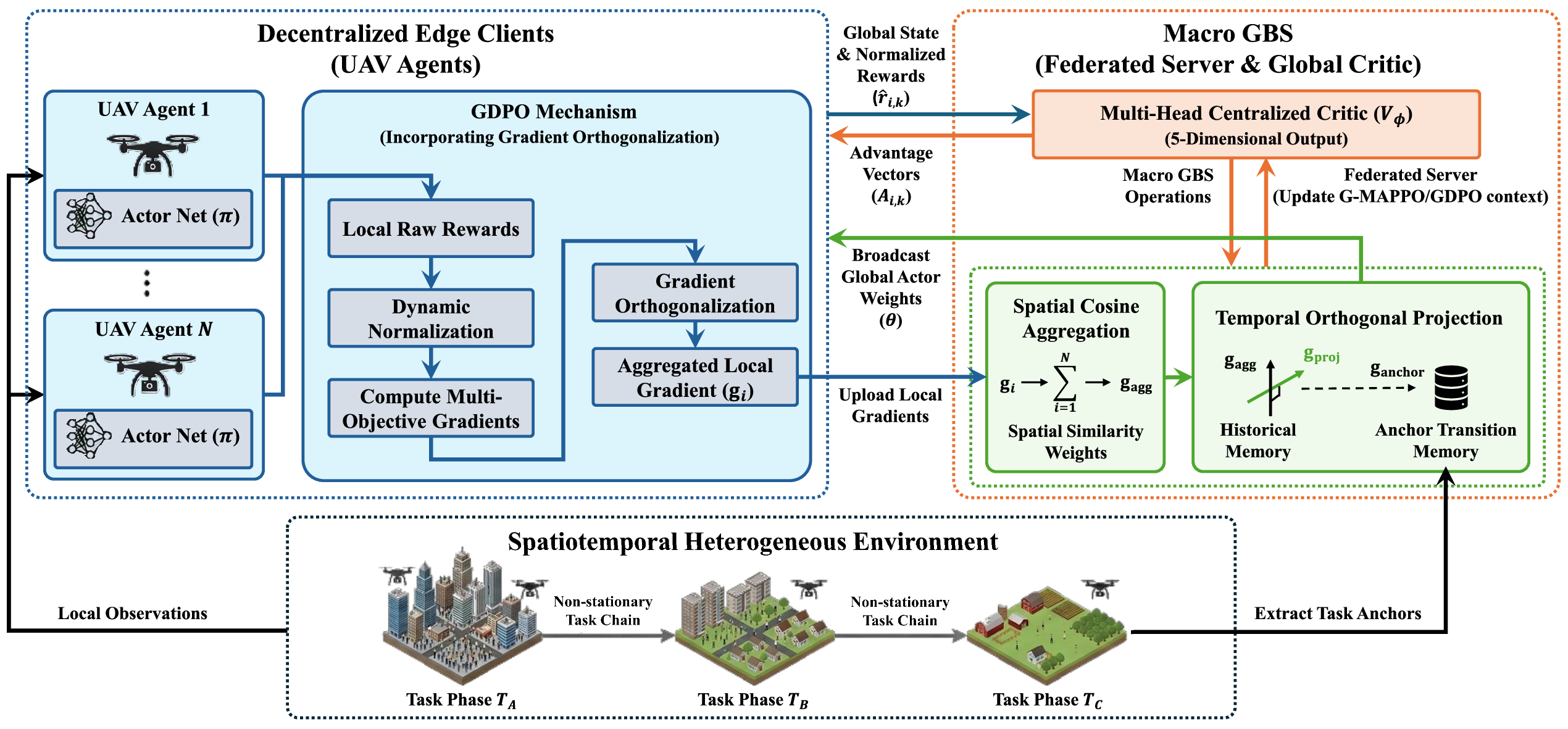}
    \caption{Schematic overview of the proposed SCFL framework under a federated CTDE paradigm. The architecture decouples the learning process into decentralized edge clients (UAV agents) and a macro GBS acting as the federated server and global critic. To resolve multi-objective conflicts and environmental non-stationarity, SCFL integrates a three-stage geometric alignment mechanism: local GDPO with gradient orthogonalization at the edge, spatial cosine aggregation at the server to mitigate non-IID client drifts, and temporal orthogonal projection utilizing an anchor transition memory. This collaborative architecture effectively preserves historical knowledge and prevents catastrophic forgetting across the continuous spatiotemporal task chain.}
    \label{fig:scfl_architecture}
    \vspace{-1em}
\end{figure*}

To address non-stationary environments and the high backhaul overhead of centralized training, we propose the SCFL framework. This framework integrates the \textit{Centralized Training with Decentralized Execution} (CTDE) paradigm into a federated edge intelligence architecture. The macro GBS acts as a federated parameter server and global critic, while the UAVs function as distributed edge actors. The schematic overview of the proposed SCFL framework is shown in Fig.~\ref{fig:scfl_architecture}.

The architecture is structured around two primary neural components:
\begin{itemize}
    \item \textbf{Decentralized Actor} ($\pi_\theta$): Each UAV agent employs a local actor network to map the filtered local observation $o_i(t)$ to a probability distribution over the 3D action space. Utilizing only local sensory data during execution keeps the control loop computationally lightweight and robust to communication delays.
    \item \textbf{Centralized Critic} ($V_\phi$): To mitigate instability from concurrent multi-agent learning, a centralized critic evaluates joint actions using the global state $s(t)$ during the federated training phase. Specifically, the critic employs a multi-head architecture yielding a $|\mathcal{K}|$-dimensional output corresponding to the heterogeneous reward components in $\mathcal{K}$, providing multi-dimensional advantage baselines for the decentralized actors.
\end{itemize}

Rather than transmitting raw trajectories, each UAV computes local actor gradients based on advantage signals. The GBS aggregates these gradients to update the global actor model. To address multi-objective conflicts, spatial heterogeneity, and temporal catastrophic forgetting during this exchange, we introduce a geometric alignment mechanism, detailed in subsequent sections.

\subsection{Decentralized POMDP Formulation}

The sequential decision-making process within the dynamic aerial-ground network is formalized as a \textit{Decentralized Partially Observable Markov Decision Process} (Dec-POMDP), defined by the tuple $\langle \mathcal{U}, \mathcal{S}, \mathcal{A}, \mathcal{O}, P, R, \gamma \rangle$.

\subsubsection{Global State Space}
The global state $s(t) \in \mathcal{S}$ encapsulates the joint configuration of the UAV swarm and the ground user distribution at time step $t$:
\begin{equation}
    s(t) = \left\{ \bigcup_{i \in \mathcal{U}} \mathbf{q}_i(t), \bigcup_{i \in \mathcal{U}} \mathbf{v}_i(t), \Phi(\mathbf{p}, t) \right\},
\end{equation}
where $\Phi(\mathbf{p}, t)$ is the instantaneous user distribution from the hybrid mobility model. The centralized critic exclusively accesses $s(t)$ during training for value estimation.

\subsubsection{Observation Space}
Due to onboard sensing constraints, each UAV $i$ perceives a localized subset of the environment. The local observation $o_i(t) \in \mathcal{O}$ supports distributed collision avoidance and service optimization:
\begin{equation}
    o_i(t) = \{ \mathbf{q}_i(t), \mathbf{v}_i(t), \mathcal{I}_{\mathrm{neigh}}^i(t), \mathcal{M}_{\mathrm{cov}}^i(t) \}.
\end{equation}
Here, $\mathcal{I}_{\mathrm{neigh}}^i(t)$ encodes spatial relationships with neighboring agents for safety, and $\mathcal{M}_{\mathrm{cov}}^i(t)$ captures the channel conditions of users within the coverage radius.

\subsubsection{Action Space}
We define a discrete action space $\mathcal{A}$. At each timestep $t$, agent $i$ selects an action $a_i(t)$:
\begin{equation}
    a_i(t) \in \{ \Delta x_{\pm}, \Delta y_{\pm}, \Delta h_{\pm}, \mathrm{Hover} \}.
\end{equation}
Horizontal commands enable tracking of user hotspots, while vertical commands $\Delta h_{\pm}$ adjust the coverage footprint and mitigate co-channel interference.

\subsection{Heterogeneous Reward Composition}
\label{subsec:reward_composition}

UAV network orchestration requires optimizing multiple conflicting indicators. The raw reward vector for each agent $i$ comprises five metrics:
\begin{equation}
    \mathbf{r}_i^{\mathrm{raw}}(t) = [ r_{\mathrm{EE}}, r_{\mathrm{Fair}}, r_{\mathrm{Load}}, r_{\mathrm{Cov}}, r_{\mathrm{QoS}} ]^T.
\end{equation}
Specifically, $r_{\mathrm{EE}}$ represents energy efficiency, incorporating the non-linear propulsion energy $P_{\mathrm{fly}}(V_i(t))$ from Section \ref{sec:system_model}. The components $r_{\mathrm{Fair}}$ and $r_{\mathrm{Load}}$ measure user rate equity and fleet load balancing via \textit{Jain's fairness index} (JFI)~\cite{JFI_1984}. The term $r_{\mathrm{Cov}}$ indicates spatial service reliability, while $r_{\mathrm{QoS}}$ penalizes worst-case user rates. These heterogeneous metrics possess distinct magnitudes and conflicting directions, which are resolved through the proposed alignment mechanism. It is important to note that these components serve as local (micro-level) instantaneous rewards to guide decentralized agent policies; the corresponding global (system-level) performance metrics used for macroscopic evaluation are explicitly defined in the simulation section.

\subsection{Three-Stage Geometric Alignment Mechanism}

The SCFL framework integrates a three-stage geometric alignment mechanism into the \textit{Group-Decoupled Multi-Agent Proximal Policy Optimization} (G-MAPPO) algorithm. The detailed algorithm of SCFL is presented in Algorithm~\ref{alg:gmappo}. This mechanism decouples policy optimization into three alignment dimensions:
\begin{itemize}
    \item \textbf{Local objective alignment (Stage 1)}: At the agent level, \textit{Group-Decoupled Policy Optimization} (GDPO) neutralizes scale disparities and resolves gradient conflicts among heterogeneous metrics before transmission.
    \item \textbf{Spatial agent alignment (Stage 2)}: At the server level, spatial cosine aggregation mitigates client drift caused by non-IID spatial distributions across edge nodes.
    \item \textbf{Temporal task alignment (Stage 3)}: Across the task chain, the GBS employs temporal orthogonal gradient projection to prevent destructive interference between new updates and historical knowledge, mitigating catastrophic forgetting.
\end{itemize}

Separating these challenges provides a scalable solution. The mathematical operations are detailed below.

\subsubsection{Stage 1: Local Group-Decoupled Policy Optimization and Global Shaping}
At the client level, gradient dominance occurs when the scale of one reward suppresses others. Dynamic z-score normalization neutralizes these disparities. Let $\mu_k$ and $\nu_k$ denote the running mean and standard deviation of the $k$-th objective. The normalized reward $\hat{r}_{i,k}(t)$ for agent $i$ is:
\begin{equation}
    \hat{r}_{i,k}(t) = \frac{r_{i,k}^{\mathrm{raw}}(t) - \mu_k}{\nu_k + \epsilon}.
\end{equation}

These normalized objectives serve two purposes. First, decentralized actors use them to compute objective-wise local gradients $\mathbf{g}_{i,k} = \nabla_\theta \mathcal{L}_{i,k}(\theta)$ for each objective $k \in \mathcal{K}$. To eliminate destructive multi-objective conflicts before server transmission, we employ a pairwise gradient projection mechanism (inspired by projected conflicting gradients). Specifically, when evaluating objective gradients, the objectives are processed in a randomly permuted order at each update step to prevent order bias. If a directional conflict occurs between two objectives ($i.e.$, $\mathbf{g}_{i,k} \cdot \mathbf{g}_{i,j} < 0$), the gradient $\mathbf{g}_{i,k}$ is projected onto the normal plane of $\mathbf{g}_{i,j}$:
\begin{equation}\label{eq:local_moo_proj}
    \mathbf{g}_{i,k} \leftarrow \mathbf{g}_{i,k} - \frac{\mathbf{g}_{i,k} \cdot \mathbf{g}_{i,j}}{\|\mathbf{g}_{i,j}\|^2 + \epsilon} \mathbf{g}_{i,j},
\end{equation}
where $\epsilon$ is a small numerical stabilizer, ensuring the projection is practically equivalent to exact orthogonalization when $\|\mathbf{g}_{i,j}\|^2 \gg \epsilon$. This yields a consolidated local gradient for agent $i$: $\mathbf{g}_i = \sum_{k \in \mathcal{K}} \mathbf{g}_{i,k}$.

Second, to explicitly address the multi-objective nature without resorting to linear scalarization, we establish a strict information flow: local raw rewards are first normalized to $\hat{r}_{i,k}$, which are then utilized by the centralized multi-head critic $V_\phi$. By aligning the numerical variance across heterogeneous rewards, this preceding dynamic $z$-score normalization stabilizes the critic's shared hidden layers. Consequently, the critic reliably outputs a $|\mathcal{K}|$-dimensional value vector $\mathbf{V}_\phi(s)$, enabling the computation of an objective-wise generalized advantage estimation (GAE) vector, $A_{i,k}$, for each agent $i$ and objective $k \in \mathcal{K}$. This multi-dimensional advantage signal directly drives the decentralized actors to compute the objective-wise local gradients $\mathbf{g}_{i,k} = \nabla_\theta \log \pi_\theta(a_i|o_i) A_{i,k}$, which are subsequently orthogonalized via~\eqref{eq:local_moo_proj} to form the consolidated local update.

\begin{algorithm2e}[!t]
\small
\SetAlgoLined
\caption{Spatiotemporal Continual Federated Learning (SCFL) Framework}
\label{alg:gmappo}
\KwIn{Task sequence $\mathcal{T}$, total episodes $E$, steps per episode $T_{\mathrm{hor}}$}
\KwInit{Global actor $\pi_\theta$, centralized critic $V_\phi$; Anchor memory $\mathcal{M}_{\mathrm{anchor}} = \emptyset$; GDPO statistics $\{\mu_k, \nu_k\}$}
\For{task $T \in \mathcal{T}$}{
    \For{episode $e = 1$ to $E$}{
        
        \tcc{Decentralized Trajectory Sampling}
        \For{$t = 1$ to $T_{\mathrm{hor}}$}{
            \ForEach{UAV $i \in \mathcal{U}$}{
                Sample $a_i(t) \sim \pi_{\theta}(\cdot | o_i(t))$\;
                Observe $o_i(t+1)$ and raw rewards $\mathbf{r}_i^{\mathrm{raw}}(t)$\;
                Store transition $(o_i(t), a_i(t), \mathbf{r}_i^{\mathrm{raw}}(t), o_i(t+1))$ locally\;
            }
        }
        
        \tcc{Reward Normalization \& Advantage Estimation}
        \ForEach{UAV $i \in \mathcal{U}$}{
            Normalize local raw rewards to $\hat{r}_{i,k}(t)$ using global statistics\;
        }
        \text{At GBS:}\;
        Update centralized multi-head critic $V_\phi$ by minimizing vector value loss on $s(t)$ using $\hat{r}_{i,k}(t)$\;
        Evaluate multi-dimensional GAE vector $A_{i,k}$ for all agents and objectives\;
        
        \tcc{Stage 1: Local Objective Alignment}
        \ForEach{UAV $i \in \mathcal{U}$}{
            Compute objective-wise local gradients $\mathbf{g}_{i,k}$\;
            Resolve multi-objective conflicts via~\eqref{eq:local_moo_proj} to yield consolidated $\mathbf{g}_i$\;
            Upload local gradient $\mathbf{g}_i$ to the GBS\;
        }
        
        \tcc{Stage 2: Spatial Agent Alignment}
        \text{At GBS:}\;
        Compute average consensus gradient $\mathbf{g}_{\mathrm{avg}} = \frac{1}{N} \sum_{i=1}^N \mathbf{g}_i$\;
        Calculate similarity weights $\omega_i$ and aggregate $\mathbf{g}_{\mathrm{agg}} = \sum_{i=1}^{N} \tilde{\omega}_i \mathbf{g}_i$\;
        
        \tcc{Stage 3: Temporal Task Alignment}
        Initialize $\mathbf{g}_{\mathrm{proj}} \leftarrow \mathbf{g}_{\mathrm{agg}}$\;
        \ForEach{historical gradient $\mathbf{g}_{\mathrm{anchor}} \in \mathcal{M}_{\mathrm{anchor}}$}{
            \If{$\mathbf{g}_{\mathrm{proj}} \cdot \mathbf{g}_{\mathrm{anchor}} < 0$}{
                $\mathbf{g}_{\mathrm{proj}} \leftarrow \mathbf{g}_{\mathrm{proj}} - \frac{\mathbf{g}_{\mathrm{proj}} \cdot \mathbf{g}_{\mathrm{anchor}}}{\|\mathbf{g}_{\mathrm{anchor}}\|^2} \mathbf{g}_{\mathrm{anchor}}$\;
            }
        }
        Update global actor $\theta \leftarrow \theta - \alpha_{\mathrm{lr}} \mathbf{g}_{\mathrm{proj}}$ and broadcast to all UAVs\;
    }
    
    \tcc{Memory Update}
    Extract anchor gradient $\mathbf{g}_{\mathrm{anchor}}$ on task $T$ and store into $\mathcal{M}_{\mathrm{anchor}}$\;
}
\end{algorithm2e}

\subsubsection{Stage 2: Spatial Federated Cosine Aggregation}
The spatial user distribution in aerial networks is highly heterogeneous, causing a non-IID data problem. For example, a UAV over a dense hotspot experiences different traffic and interference than one in a sparse area. Standard federated averaging is vulnerable to client drift under these conditions, as divergent local gradients can destabilize the global model.

To align spatial knowledge and mitigate drift, the GBS performs adaptive cosine-based aggregation. First, it computes an unweighted average gradient representing the general consensus:
\begin{equation}
    \mathbf{g}_{\mathrm{avg}} = \frac{1}{N} \sum_{i=1}^N \mathbf{g}_i.
\end{equation}
Each local gradient is then evaluated by its cosine similarity to this consensus. A dynamic weight $\omega_i$ is assigned to agent $i$:
\begin{equation}
    \omega_i = \max \left( \delta, \frac{\mathbf{g}_i \cdot \mathbf{g}_{\mathrm{avg}}}{\|\mathbf{g}_i\| \|\mathbf{g}_{\mathrm{avg}}\|} \right),
\end{equation}
where $\delta = 0.01$ is a minimum threshold preventing the elimination of unique divergent experiences. The spatially aggregated gradient $\mathbf{g}_{\mathrm{agg}}$ is the normalized weighted sum of all local gradients:
\begin{equation}
    \mathbf{g}_{\mathrm{agg}} = \sum_{i=1}^{N} \tilde{\omega}_i \mathbf{g}_i, \quad \text{where } \tilde{\omega}_i = \frac{\omega_i}{\sum_{j=1}^{N} \omega_j}.
\end{equation}
This mechanism attenuates the contribution of highly divergent client gradients from non-IID environments. By down-weighting conflicting gradients, the GBS maintains a stable evolutionary trajectory.

\subsubsection{Stage 3: Temporal Orthogonal Gradient Projection}
To prevent catastrophic forgetting across tasks, the GBS maintains anchor experiences from previous tasks. Let $\mathbf{g}_{\mathrm{anchor}}$ denote the reference anchor gradient evaluated on historical transitions. When a directional conflict is detected ($\mathbf{g}_{\mathrm{agg}} \cdot \mathbf{g}_{\mathrm{anchor}} < 0$), the GBS projects the aggregated gradient onto the normal plane of $\mathbf{g}_{\mathrm{anchor}}$:
\begin{equation} \label{eq:projection}
    \mathbf{g}_{\mathrm{proj}} = \mathbf{g}_{\mathrm{agg}} - \frac{\mathbf{g}_{\mathrm{agg}} \cdot \mathbf{g}_{\mathrm{anchor}}}{\|\mathbf{g}_{\mathrm{anchor}}\|^2} \mathbf{g}_{\mathrm{anchor}}.
\end{equation}
If no conflict exists ($\mathbf{g}_{\mathrm{agg}} \cdot \mathbf{g}_{\mathrm{anchor}} \geq 0$), the consensus direction is preserved such that $\mathbf{g}_{\mathrm{proj}} = \mathbf{g}_{\mathrm{agg}}$. This operation orthogonalizes the update direction against prior tasks, preserving historical knowledge without raw data replay.

\subsection{Computational Complexity Analysis}

The SCFL framework partitions computational workload for efficiency. During decentralized inference, each UAV executes a forward pass through its actor network. For an architecture with $L_a$ layers and maximum hidden dimension $D_h$, local time complexity per step is bounded by $O(L_a D_h^2)$. Utilizing only local observations makes execution independent of swarm size $N$ and user density $M$, satisfying the latency requirements of aerial navigation.

During federated training, the burden is distributed between clients and the GBS. At the UAV level, computing gradients for $|\mathcal{K}|$ objectives over a mini-batch of size $B_{\mathrm{size}}$ for $E_{\mathrm{ppo}}$ epochs incurs a complexity of $O(|\mathcal{K}| E_{\mathrm{ppo}} B_{\mathrm{size}} L_a D_h^2)$. Local gradient orthogonalization adds an overhead of $O(|\mathcal{K}|^2 |\theta_a|)$, where $|\theta_a|$ is the parameter count in the actor network. At the GBS level, updating the critic network requires $O(E_{\mathrm{ppo}} B_{\mathrm{size}} L_c D_h^2)$ operations, with $L_c$ denoting critic layers. Processing spatial cosine aggregation across $N$ agents and temporal projection against $M_{\mathrm{hist}}$ tasks requires $O(N |\theta_a|)$ and $O(M_{\mathrm{hist}} |\theta_a|)$ operations, respectively. 

Since $|\mathcal{K}|$, $N$, and $M_{\mathrm{hist}}$ are typically small constants, the geometric projection layers introduce negligible overhead relative to standard backpropagation. Ultimately, by avoiding the repeated transmission and storage of raw historical trajectories or experience replay buffers across bandwidth-constrained wireless links, this paradigm trades local edge computation for data privacy and operational efficiency. Specifically, edge clients only upload model-sized actor-gradient vectors every federated aggregation interval (e.g., 50 steps), satisfying the practical constraints of continual aerial swarm orchestration.

\section{Theoretical Analysis of Knowledge Retention}
\label{sec:theoretical_analysis}

This section provides a mathematical analysis of knowledge retention in the G-MAPPO algorithm under non-stationary federated environments through the lens of \textit{Multi-Objective Optimization} (MOO). It demonstrates that the temporal gradient projection mechanism prevents the acquisition of new spatiotemporal policies from interfering with previously consolidated knowledge at the global server level.

Knowledge interference occurs when the UAV swarm transitions between distinct geographical scenarios. In the federated context, global parameter updates represent a trajectory search between task-specific manifolds driven by aggregated local experiences.

\begin{definition}[Sequential Task Interference]
    Consider a sequential task chain where the global model first optimizes a policy for a task $T_{\mathrm{prev}}$, and subsequently adapts to a new task $T_{\mathrm{new}}$. Let $\mathbf{g}_{\mathrm{anchor}}$ denote the stored reference anchor gradient extracted at the end of $T_{\mathrm{prev}}$. We assume $\mathbf{g}_{\mathrm{anchor}}$ serves as a local surrogate for the historical loss gradient under slowly varying gradient fields. For the new task, the GBS gathers local gradients and computes the spatially aggregated gradient $\mathbf{g}_{\mathrm{agg}}$.
\end{definition}

In standard federated learning updates, global parameters move along the negative direction of the aggregated gradient for the current task, which may shift the model away from previous optima. A \textit{first-order Taylor expansion} is utilized to approximate the trend of the previous loss function and define the mathematical boundary of catastrophic forgetting.

\begin{proposition}[Catastrophic Forgetting Condition]
    When the global parameters are updated according to the new aggregated gradient ($\theta \leftarrow \theta - \alpha_{\mathrm{lr}} \mathbf{g}_{\mathrm{agg}}$), the change in the previous loss function is approximated by $\mathcal{L}_{\mathrm{prev}}(\theta - \alpha_{\mathrm{lr}} \mathbf{g}_{\mathrm{agg}}) \approx \mathcal{L}_{\mathrm{prev}}(\theta) - \alpha_{\mathrm{lr}} (\mathbf{g}_{\mathrm{anchor}} \cdot \mathbf{g}_{\mathrm{agg}})$, where $\alpha_{\mathrm{lr}} > 0$ denotes the global learning rate. Consequently, catastrophic forgetting is triggered if and only if $\mathbf{g}_{\mathrm{anchor}} \cdot \mathbf{g}_{\mathrm{agg}} < 0$. This inequality indicates that the consensus update direction for the new task conflicts with the descent direction of the prior task, leading to a localized increase in the previous loss.
\end{proposition}

This directional conflict occurs due to a negative correlation between the aggregated and historical gradients. To address this, G-MAPPO performs a geometric correction at the GBS before the global broadcast. By projecting the conflicting aggregated gradient onto the normal plane of the previous gradient, a conflict-free, Pareto-compatible direction is obtained to explore new strategies while maintaining historical performance.

\begin{definition}[Orthogonal Gradient Projection]
    To eliminate destructive interference with prior task manifolds, the GBS computes the projected gradient $\mathbf{g}_{\mathrm{proj}}$ according to the geometric projection operator formulated in~\eqref{eq:projection}.
\end{definition}

Following the principle of gradient surgery, the projection operator makes the update orthogonal to the stored historical gradient, corresponding to a tangent direction of the local level set of the historical loss under the stated surrogate assumption. Based on this construction, the following theorem guarantees that the federated system maintains historical performance during environmental transitions.

\begin{proposition}[First-Order Directional Protection]
    Under the gradient projection update rule $\theta \leftarrow \theta - \alpha_{\mathrm{lr}} \mathbf{g}_{\mathrm{proj}}$, the inner product between the projected gradient and the historical anchor gradient satisfies $\mathbf{g}_{\mathrm{anchor}} \cdot \mathbf{g}_{\mathrm{proj}} = 0$. Consequently, the update is orthogonal to the stored historical anchor direction, providing first-order directional protection against historical interference, such that $\mathcal{L}_{\mathrm{prev}}(\theta - \alpha_{\mathrm{lr}} \mathbf{g}_{\mathrm{proj}}) \approx \mathcal{L}_{\mathrm{prev}}(\theta)$.
\end{proposition}

\begin{proof}
    By substituting the projected gradient defined in \eqref{eq:projection} into the Taylor expansion of the previous loss, the inner product of the parameter updates is calculated as:
    \begin{align}
        \mathbf{g}_{\mathrm{anchor}} \cdot \mathbf{g}_{\mathrm{proj}} &= \mathbf{g}_{\mathrm{anchor}} \cdot \left( \mathbf{g}_{\mathrm{agg}} - \frac{\mathbf{g}_{\mathrm{agg}} \cdot \mathbf{g}_{\mathrm{anchor}}}{\|\mathbf{g}_{\mathrm{anchor}}\|^2} \mathbf{g}_{\mathrm{anchor}} \right) \notag \\
        &= \mathbf{g}_{\mathrm{anchor}} \cdot \mathbf{g}_{\mathrm{agg}} - \frac{\mathbf{g}_{\mathrm{agg}} \cdot \mathbf{g}_{\mathrm{anchor}}}{\|\mathbf{g}_{\mathrm{anchor}}\|^2} (\mathbf{g}_{\mathrm{anchor}} \cdot \mathbf{g}_{\mathrm{anchor}}) \notag \\
        &= \mathbf{g}_{\mathrm{anchor}} \cdot \mathbf{g}_{\mathrm{agg}} - (\mathbf{g}_{\mathrm{agg}} \cdot \mathbf{g}_{\mathrm{anchor}}) = 0.
    \end{align}
    Since $\mathbf{g}_{\mathrm{anchor}} \cdot \mathbf{g}_{\mathrm{proj}} = 0$, the parameter update trajectory is restricted to the tangent space of the local level set of the previous task's loss at $\theta$. Therefore, the first-order loss variation is eliminated, preventing immediate directional interference during sequential task transitions. Note that when maintaining multiple historical anchors, a sequential projection scheme is employed to suppress directional conflicts across task phases.
\end{proof}

While the projection mechanism addresses temporal directional conflicts at the global level, multi-objective learning stability depends on the numerical scale of local gradients. The framework achieves equilibrium by integrating local normalization, spatial aggregation, and temporal projection.

\begin{remark}[Stability-Plasticity Synergy in SCFL]
    The SCFL framework relies on three algorithmic stages. Local group-decoupled normalization balances gradient magnitudes across heterogeneous rewards. Spatial cosine aggregation mitigates non-IID client drift. Temporal gradient projection provides a geometric constraint for directional non-interference. Together, these mechanisms address the stability-plasticity dilemma in non-stationary aerial networks, enabling the exploration of new policies (plasticity) while preserving historical knowledge (stability).
\end{remark}

\section{Simulation Results and Analysis}
\label{sec:simulation_results}

This section systematically evaluates the performance of the proposed SCFL framework across three distinct dimensions: 1) dynamic adaptability and policy stabilization during sequential spatiotemporal task transitions under a fixed density ($M=100$); 2) scalability and congestion stress analysis across varying user densities ($M \in [40, 140]$); and 3) longitudinal self-degradation and mechanism ablation to quantify memory retention and reveal the physical boundaries of projection constraints.

\subsection{Simulation Setup and Performance Evaluation Metrics}
\label{subsec:setup}

The simulation parameters are summarized in Table~\ref{tab:parameters}. We consider a mobile edge network area of $2 \times 2$ km$^2$. To assess the resilience against non-stationary concept drifts, the simulation environment undergoes sequential phase transitions among three distinct spatiotemporal regimes:
\begin{itemize}
    \item \textit{Crowded Urban Phase}: Characterized by a high user density of $M=140$ with highly clustered distributions. The primary challenge involves managing co-channel interference and capacity offloading for the GBS.
    \item \textit{Suburban Phase}: Features moderate user density ($M=80 \sim 100$) with semi-clustered distributions. It requires the swarm to balance spectral efficiency with regional coverage.
    \item \textit{Rural Phase}: Marked by a low user density of $M=40$ with sparse distributions. The objective shifts toward maximizing coverage probability and extending the service footprint.
\end{itemize}

The default swarm size is $N=4$ for the service area. This sparse deployment forces agents to dynamically prioritize mission objectives, serving as a benchmark for evaluating the resilience of the SCFL framework against catastrophic forgetting compared to traditional federated baselines.

\begin{table}[t]
\caption{Simulation Parameters and SCFL Hyperparameters}
\label{tab:parameters}
\centering
\begin{tabular}{@{}ll@{}}
\toprule
\textbf{Parameter} & \textbf{Value} \\ \midrule
\multicolumn{2}{c}{\textit{Spatiotemporal Environment Settings}} \\ \midrule
Service Area & $2 \times 2$ km$^2$ \\
Number of UAVs ($N$) & 4 \\
User Density ($M$) & 40, 60, 80, 100, 120, 140 \\
Mobility Model & Hybrid RPGM \& Gauss-Markov \\
Velocity Memory ($\alpha_m$) / Noise ($\sigma_z$) & 0.8 / 1.0 m/s \\
UAV Altitude Range ($[H_{\min}, H_{\max}]$) & $[80, 120]$ m \\
Episodes per Phase / Independent Runs & 100 / 10 runs \\
Steps per Episode ($T_{\mathrm{hor}}$) & 200 \\ 
\midrule
\multicolumn{2}{c}{\textit{Heterogeneous Communication Model}} \\ \midrule
Carrier Freq. ($f_c$) / Bandwidth ($B$) & 2 GHz / 20 MHz \\
Noise Power Density ($N_0$) / Power ($\sigma^2$) & -174 dBm/Hz \\
Transmit Power ($P_{\mathrm{UAV}}$, $P_{\mathrm{GBS}}$) & 23, 43 dBm \\
Antenna Gain ($G_{\mathrm{UAV}}^{\mathrm{ant}}$, $G_{\mathrm{GBS}}^{\mathrm{ant}}$) & 2, 15 dBi \\
Path Loss Model & Prob. LoS/NLoS~\cite{al2014modeling} \\ \midrule
\multicolumn{2}{c}{\textit{UAV Aerodynamic \& Propulsion Parameters}} \\ \midrule
Profile ($P_0$) / Induced Power ($P_{\mathrm{ind}}$) & 79.856 W / 88.628 W \\
Rotor Tip Speed ($U_{\mathrm{tip}}$) & 120.0 m/s \\
Mean Induced Velocity ($v_0$) & 4.03 m/s \\
Fuselage Drag Ratio ($d_{\mathrm{fus}}$) & 0.6 \\
Air Density ($\rho$) & 1.225 kg/m$^3$ \\
Solidity ($s_{\mathrm{rotor}}$) / Disc Area ($A$) & 0.05 / 0.503 m$^2$ \\ \midrule
\multicolumn{2}{c}{\textit{Federated Learning \& Algorithm Hyperparameters}} \\ \midrule
Learning Rate (Actor, Critic) & $10^{-4}$, $10^{-3}$ \\
Hidden Layers $D_h$ (Actor, Critic) & $[128, 128]$ \\
Critic Architecture & Multi-Head ($|\mathcal{K}| = 5$) \\
Discount ($\gamma$) / GAE ($\lambda$) & 0.99 / 0.95 \\
Clipping Ratio ($\epsilon$) & 0.2 (Linear annealed) \\
Initial Entropy Weight & 0.05 (Exp. annealed) \\
Mini-batch Size ($B_{\mathrm{size}}$) / Epochs ($E_{\mathrm{ppo}}$) & 64 / 10 \\
Federated Aggregation Interval & 50 steps \\
Multi-Objective Weights ($\mathbf{w}$) & $[0.2, 0.3, 0.8, 1.2, 0.8]$ \\
Reward Scaling Mechanism & Dynamic $z$-score \\ \bottomrule
\end{tabular}
\vspace{1em}
\end{table}

To ensure statistical robustness, all evaluated metrics are averaged over 10 independent simulation runs under distinct random seeds. To explicitly map our macroscopic evaluation to the localized reward components defined in Section \ref{subsec:reward_composition}, we define:
\begin{itemize}
    \item \textit{Total System Throughput}: Measures the aggregate network capacity: 
    \begin{equation}\label{eq:throughput}
        C_{\mathrm{total}} = \sum_{u=1}^{M} R_u.
    \end{equation}
    
    \item \textit{User Fairness}: Evaluates equitable service distribution using \textit{Jain's fairness index} (JFI)~\cite{JFI_1984}, optimized by the localized fairness reward $r_{\mathrm{Fair}}$: 
    \begin{equation}\label{eq:jfi}
        \mathcal{J}_{\mathrm{rate}} = \frac{(\sum_{u=1}^{M} R_u)^2}{M \sum_{u=1}^{M} R_u^2}.
    \end{equation}
    
    \item \textit{Spatial Service Reliability}: The ratio of users exceeding the minimum QoS threshold $R_{\mathrm{th}} = 1$ Mbps, serving as the macroscopic metric for $r_{\mathrm{Cov}}$: 
    \begin{equation}\label{eq:coverage}
        P_{\mathrm{cov}} = \frac{1}{M} \sum_{u=1}^{M} \mathbb{I}(R_u \geq R_{\mathrm{th}}).
    \end{equation}
    
    \item \textit{Minimum QoS}: Represents the lowest achievable data rate among all users, safeguarded by the penalty component $r_{\mathrm{QoS}}$: 
    \begin{equation}\label{eq:qos}
        R_{\min} = \min_{u \in \{1, \dots, M\}} R_u.
    \end{equation}
    
    \item \textit{UAV Fleet Load Efficiency}: Assesses swarm coordination via a JFI-based load index (where $M_k$ is the number of users served by node $k$), reflecting the impact of $r_{\mathrm{Load}}$: 
    \begin{equation}\label{eq:load}
        \mathcal{J}_{\mathrm{load}} = \frac{(\sum_{k=0}^{N} M_k)^2}{(N+1) \sum_{k=0}^{N} M_k^2}.
    \end{equation}
    
    \item \textit{Total System Reward}: Tracks the unshaped environmental sum reward (aggregating components including $r_{\mathrm{EE}}$) to quantify learning convergence and stability.
\end{itemize}

\subsection{Comparative Methods and Baselines}
\label{subsec:baselines}

To isolate the contributions of the proposed framework, we compare it against four baselines. For a fair comparison, the reinforcement learning baselines are also equipped with a federated architecture (denoted by the ``Fed-'' prefix) using the GBS as a parameter server. The comparative methods are as follows:
\begin{itemize}
    \item \textbf{Proposed SCFL (Full Framework)}: Integrates all proposed mechanisms on a Fed-MAPPO backbone, including local GDPO (Stage 1), spatial cosine aggregation (Stage 2), and temporal orthogonal projection (Stage 3).
    \item \textbf{Fed-MAPPO w/ EWC (CL Baseline)~\cite{pnas2017}}: Incorporates EWC as a conventional continual learning regularizer to penalize significant parameter deviations from prior tasks, serving as a standard regularization-based benchmark. The EWC regularization coefficient was selected from $\{10, 100, 1000, 5000\}$, with $\lambda_{\mathrm{EWC}} = 10$ yielding the best overall performance.
    \item \textbf{Vanilla Fed-MAPPO (Ablation)}: Retains the Fed-MAPPO architecture but removes both local gradient orthogonalization and temporal projection, relying solely on standard FedAvg. It highlights catastrophic forgetting when alignment mechanisms are absent.
    \item \textbf{Fed-MADDPG (Conventional Baseline)}: A federated extension of MADDPG representing a standard off-policy MARL approach without advanced multi-objective conflict resolution under spatiotemporal transitions.
    \item \textbf{Static K-Means (SKM)}: A non-learning heuristic deploying UAVs at user cluster centroids. It serves as a strict spatial topological baseline, isolating the fundamental limitations of static 2D deployments without 3D altitude adjustments, co-channel interference management, or propulsion constraints.
\end{itemize}

\subsection{Task Chain Convergence and Adaptability}
\label{subsec:convergence}

\begin{figure}[!t]
    \centering
    \begin{subfigure}[b]{0.5\columnwidth}
        \includegraphics[width=\linewidth]{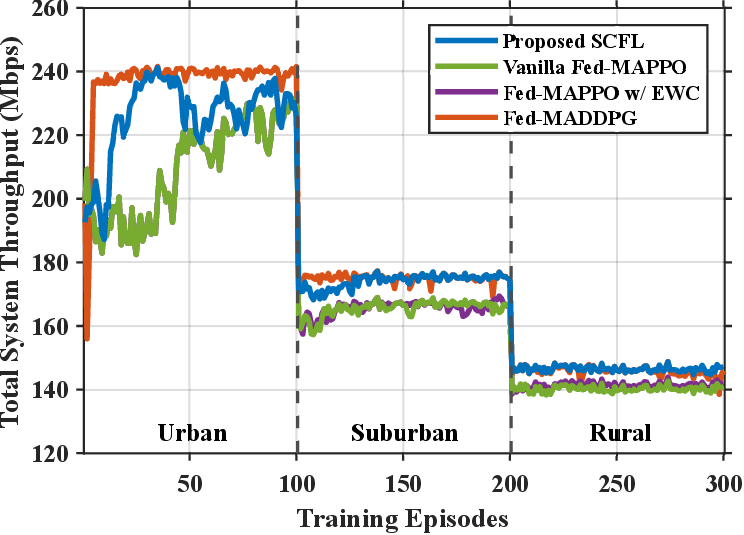}
        \caption{Total system throughput}
        \label{fig:convergence:thr}
    \end{subfigure}\hfill
    \begin{subfigure}[b]{0.5\columnwidth}
        \includegraphics[width=\linewidth]{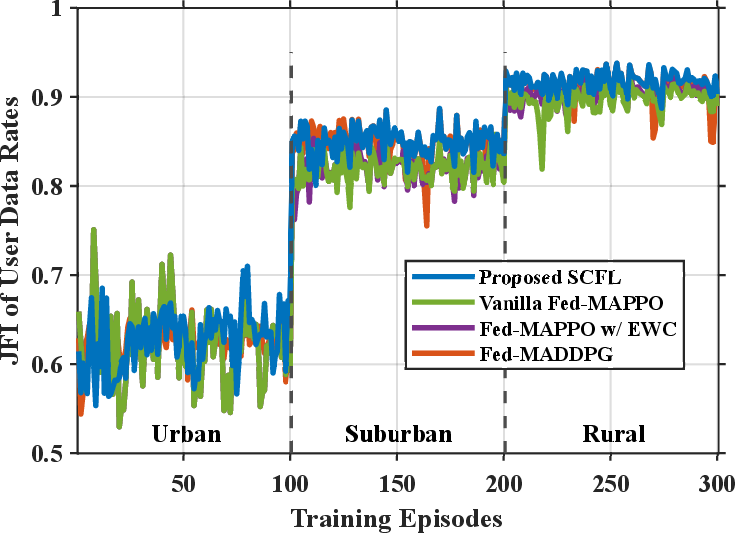}
        \caption{JFI of user data rates}
        \label{fig:convergence:JFI_rate}
    \end{subfigure}    
    
    \vspace{.2em}
    \begin{subfigure}[b]{0.5\columnwidth}
        \includegraphics[width=\linewidth]{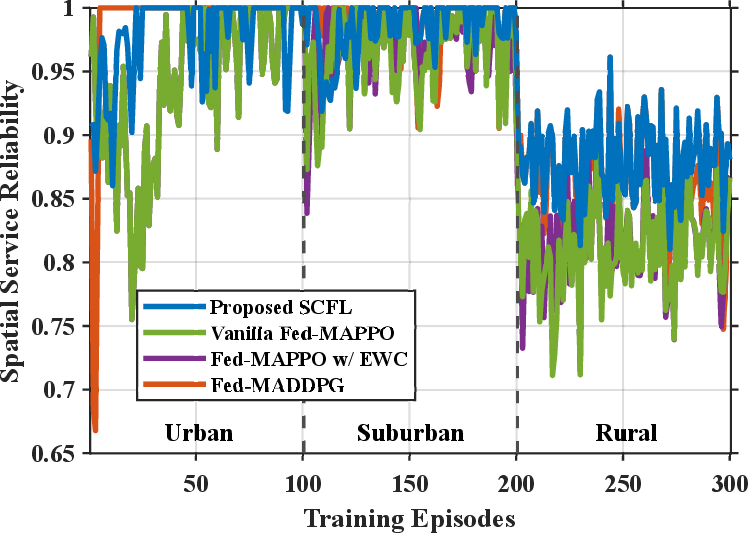}
        \caption{Spatial service reliability}
        \label{fig:convergence:coverage}
    \end{subfigure}\hfill
    \begin{subfigure}[b]{0.5\columnwidth}
        \includegraphics[width=\linewidth]{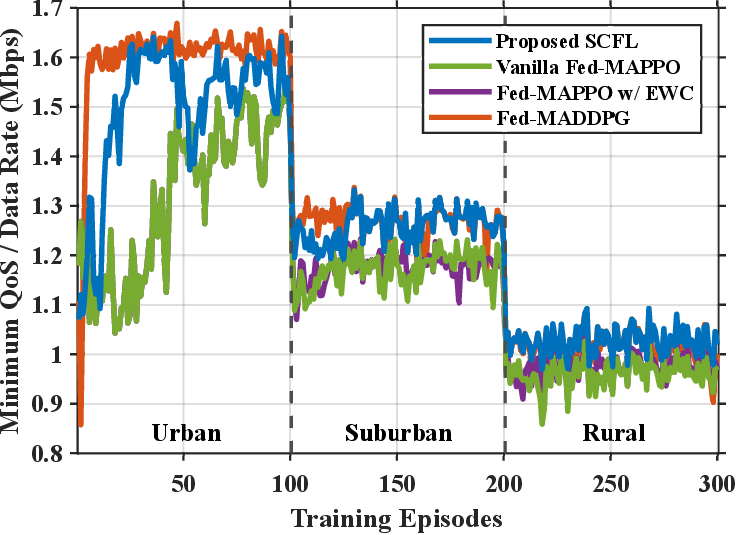}
        \caption{Minimum QoS}
        \label{fig:convergence:qos}
    \end{subfigure}
    
    \vspace{.2em}
    \begin{subfigure}[b]{0.5\columnwidth}
        \includegraphics[width=\linewidth]{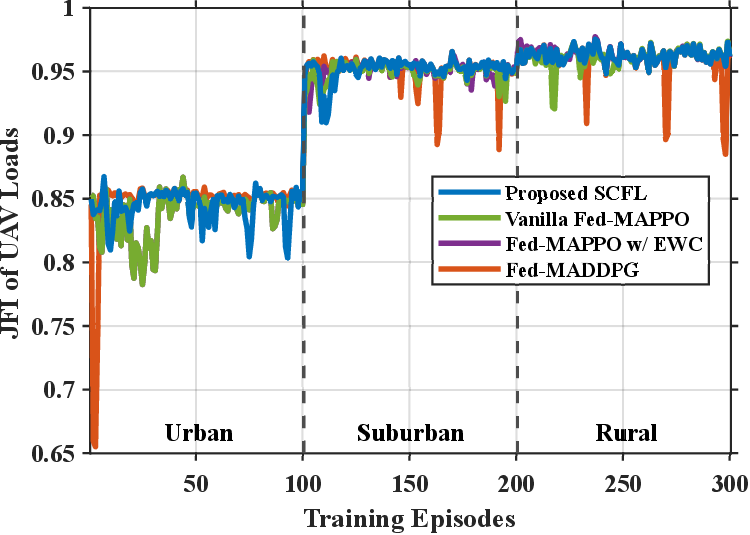}
        \caption{JFI of UAV loads}
        \label{fig:convergence:JFI_load}
    \end{subfigure}\hfill
    \begin{subfigure}[b]{0.5\columnwidth}
        \includegraphics[width=\linewidth]{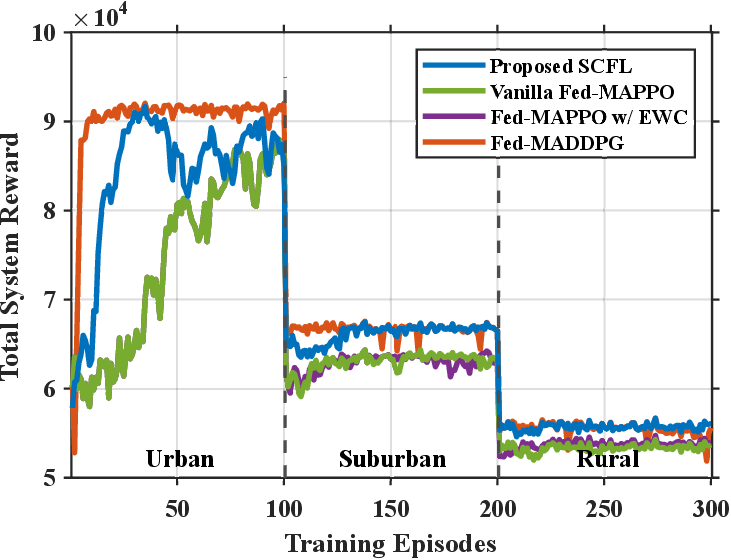}
        \caption{Total system reward}
        \label{fig:convergence:reward}
    \end{subfigure}    
    \caption{Convergence analysis of key performance metrics across the sequential spatiotemporal task chain (Urban, Suburban, and Rural) under a fixed user density of $M = 100$.}
    \label{fig:convergence}
\end{figure}

This section evaluates dynamic adaptability and policy stability under non-stationary environmental shifts across the sequential spatiotemporal task chain (Urban, Suburban, and Rural). The user density is fixed at $M=100$ to assess multi-objective coordination under constrained network resources. The convergence trajectories over 300 episodes are presented in Fig.~\ref{fig:convergence}.

During the Crowded Urban phase (Episodes 0 to 100), the clustered user distribution provides dense reward signals despite co-channel interference. Fed-MADDPG exhibits rapid convergence, reaching the peak throughput shown in Fig.~\ref{fig:convergence:thr} and a reliability ratio approaching 1.0 presented in Fig.~\ref{fig:convergence:coverage}. This performance relates to its off-policy nature and replay buffer, which reuse dense reward experiences. Conversely, on-policy algorithms require real-time trajectory sampling. Consequently, the Vanilla Fed-MAPPO baseline exhibits oscillations in the JFI of user data rates illustrated in Fig.~\ref{fig:convergence:JFI_rate} due to conflicting gradients. The proposed SCFL framework addresses this via the GDPO mechanism, orthogonalizing conflicting gradients. This enables SCFL to mitigate on-policy sampling variance, achieving comparable peak performance to Fed-MADDPG with a minimum QoS of approximately 1.6 Mbps demonstrated in Fig.~\ref{fig:convergence:qos}, while retaining policy plasticity for subsequent tasks.

Environmental transition boundaries (Episodes 100 and 200) serve as primary evaluations for dynamic adaptability. Upon shifting to the semi-clustered Suburban model at Episode 100, the Vanilla Fed-MAPPO agent experiences a drop in total system reward in Fig.~\ref{fig:convergence:reward} and minimum QoS in Fig.~\ref{fig:convergence:qos}, indicating limited spatiotemporal plasticity. Furthermore, although the traditional continual learning baseline (Fed-MAPPO w/ EWC) provides limited improvement over Vanilla Fed-MAPPO on selected metrics, stronger regularization constrains parameter plasticity and slows adaptation to new environments, as observed in Fig.~\ref{fig:convergence:thr}. In contrast, the proposed SCFL demonstrates consistent adaptability. By enforcing temporal orthogonal gradient projection (Stage 3), SCFL constrains parameter updates to a conflict-free, Pareto-compatible direction, reducing parameter overwriting during environmental shifts. Consequently, SCFL transitions steadily into the Suburban phase, maintaining higher total system rewards shown in Fig.~\ref{fig:convergence:reward} and a spatial service reliability above 0.95 presented in Fig.~\ref{fig:convergence:coverage}.

While Fed-MADDPG performs well in the dense Urban phase, its reliance on the replay buffer introduces limitations in sparse environments. During the Suburban and Rural phases, the dispersed user distribution yields sparser reward signals. As depicted in the convergence plots for JFI of UAV loads in Fig.~\ref{fig:convergence:JFI_load} and spatial service reliability in Fig.~\ref{fig:convergence:coverage}, Fed-MADDPG exhibits recurrent downward trenches below 0.90. This indicates a potential reward collapse, where off-policy agents may overfit to outdated experiences, affecting 3D trajectory coordination. Alternatively, the spatial cosine aggregation (Stage 2) in the SCFL framework mitigates non-IID client drift under sparse conditions. This mechanism facilitates synchronized fleet coordination, maintaining the JFI of UAV loads at approximately 0.96 across subsequent phases as demonstrated in Fig.~\ref{fig:convergence:JFI_load}.

\subsection{Comparative Scalability and Stress Analysis}
\label{subsec:scalability}

\begin{figure}[!t]
    \centering
    \begin{subfigure}[b]{0.5\columnwidth}
        \includegraphics[width=\linewidth]{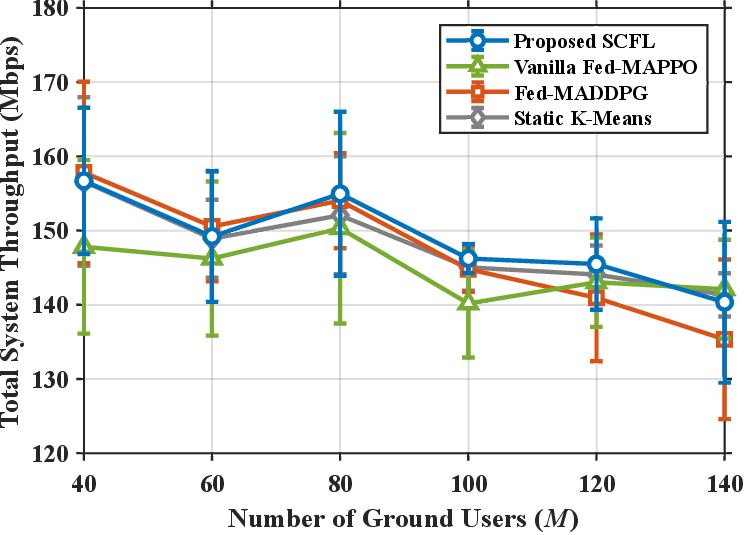}
        \caption{Total system throughput}
        \label{fig:scalability:thr}
    \end{subfigure}\hfill
    \begin{subfigure}[b]{0.5\columnwidth}
        \includegraphics[width=\linewidth]{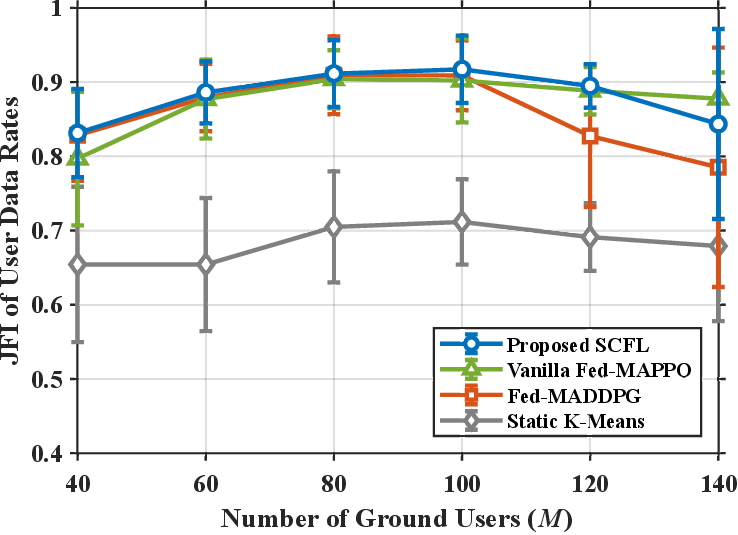}
        \caption{JFI of user data rates}
        \label{fig:scalability:JFI_rate}
    \end{subfigure}    
    
    \vspace{.2em}
    \begin{subfigure}[b]{0.5\columnwidth}
        \includegraphics[width=\linewidth]{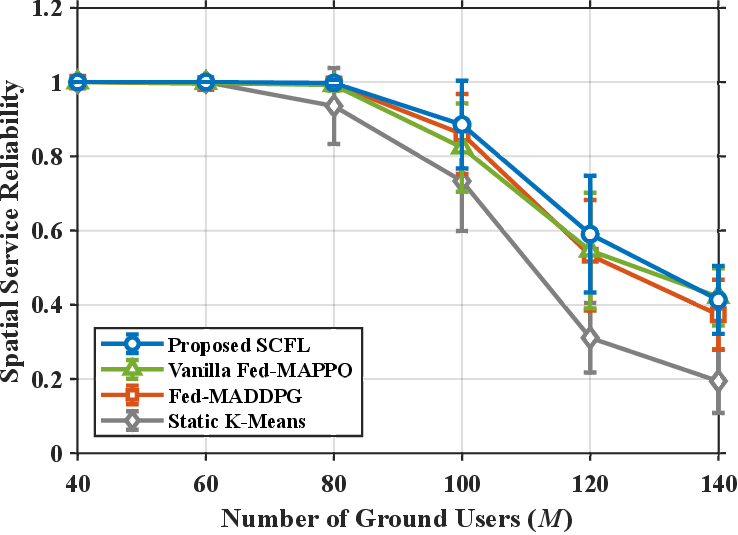}
        \caption{Spatial service reliability}
        \label{fig:scalability:coverage}
    \end{subfigure}\hfill
    \begin{subfigure}[b]{0.5\columnwidth}
        \includegraphics[width=\linewidth]{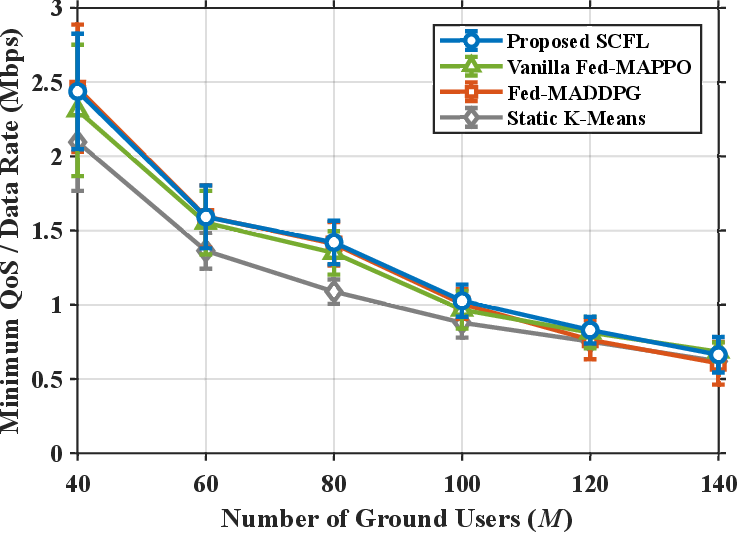}
        \caption{Minimum QoS}
        \label{fig:scalability:qos}
    \end{subfigure}

    \vspace{.2em}
    \begin{subfigure}[b]{0.5\columnwidth}
        \includegraphics[width=\linewidth]{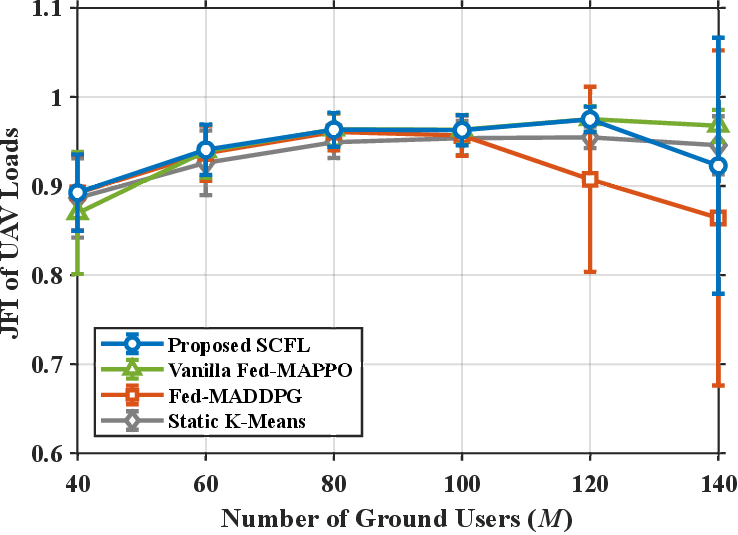}
        \caption{JFI of UAV loads}
        \label{fig:scalability:load}
    \end{subfigure}\hfill
    \begin{subfigure}[b]{0.5\columnwidth}
        \includegraphics[width=\linewidth]{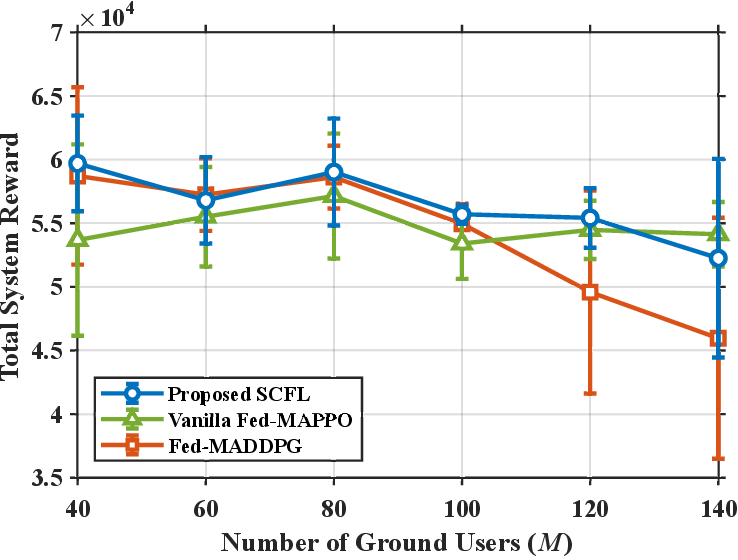}
        \caption{Total system reward}
        \label{fig:scalability:reward}
    \end{subfigure}    
    \caption{Scalability analysis of key performance metrics across varying user densities ($M \in [40, 140]$).}
    \label{fig:scalability}
\end{figure}

To evaluate system robustness under increasing network congestion, we analyze the converged performance metrics across a range of user densities ($M \in [40, 140]$). Specifically, the data points are collected from the terminal stabilization phase of the final rural scenario, reflecting the operational capability of the UAV swarm after navigating the complete spatiotemporal task sequence. The stress test results comparing the proposed SCFL framework against baseline models are illustrated in Fig.~\ref{fig:scalability}.

As shown in Fig.~\ref{fig:scalability:thr}, total system throughput exhibits a general downward trend as user density increases from $M=40$ to $M=140$, reflecting channel congestion and bandwidth exhaustion under high-load conditions. Nevertheless, the proposed SCFL framework maintains higher throughput compared to the evaluated baselines across the density spectrum. While heuristic benchmarks and off-policy models experience capacity drops, SCFL sustains its throughput by mitigating co-channel interference through spatial cosine aggregation.

Regarding service fairness illustrated in Fig.~\ref{fig:scalability:JFI_rate} and spatial service reliability presented in Fig.~\ref{fig:scalability:coverage}, the performance gap between learning-based methods and the Static K-Means benchmark widens as the network scales. Static K-Means exhibits limited adaptability to high-density clustering, resulting in a reliability ratio dropping below 0.21 at $M=140$. In contrast, SCFL maintains spatial service reliability and user fairness, performing comparably or better than Vanilla Fed-MAPPO. This observation supports that local multi-objective gradient orthogonalization mitigates the degradation of fairness parameters when network resources are bottlenecked.

Fig.~\ref{fig:scalability:qos} and Fig.~\ref{fig:scalability:load} illustrate the system's ability to maintain minimum QoS guarantees and UAV load balancing. As user counts increase, the minimum QoS requirements become harder to satisfy due to power limitations. However, SCFL secures a relatively higher margin of minimum data rates compared to baseline models. Furthermore, as depicted in Fig.~\ref{fig:scalability:load}, SCFL preserves a stable load balancing index close to 0.95 across all user scales, whereas Fed-MADDPG exhibits load fluctuations at higher densities. This indicates that the integrated spatial-temporal alignment mechanisms support stable fleet coordination and load distribution under scalability stress.

\subsection{Self-Degradation and Ablation Analysis}
\label{subsec:longitudinal_ablation_study}

\begin{figure}[!t]
    \centering
    \begin{subfigure}[b]{0.5\columnwidth}
        \includegraphics[width=\linewidth]{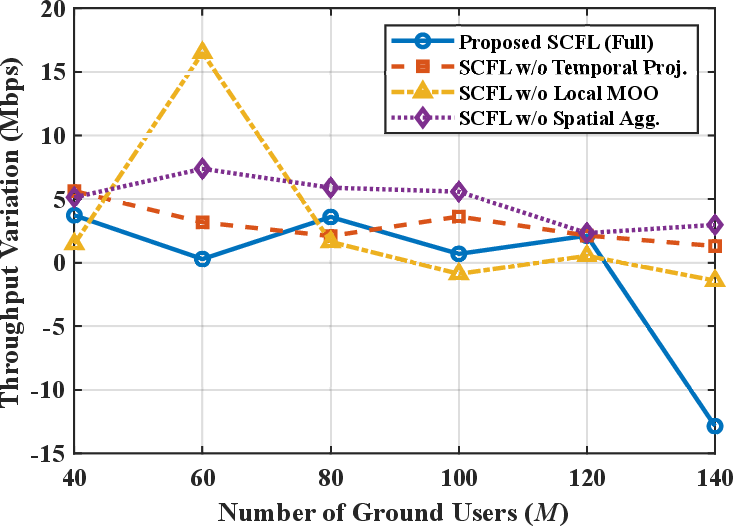}
        \caption{Variation in total system throughput}
        \label{fig:ablation:thr}
    \end{subfigure}\hfill
    \begin{subfigure}[b]{0.5\columnwidth}
        \includegraphics[width=\linewidth]{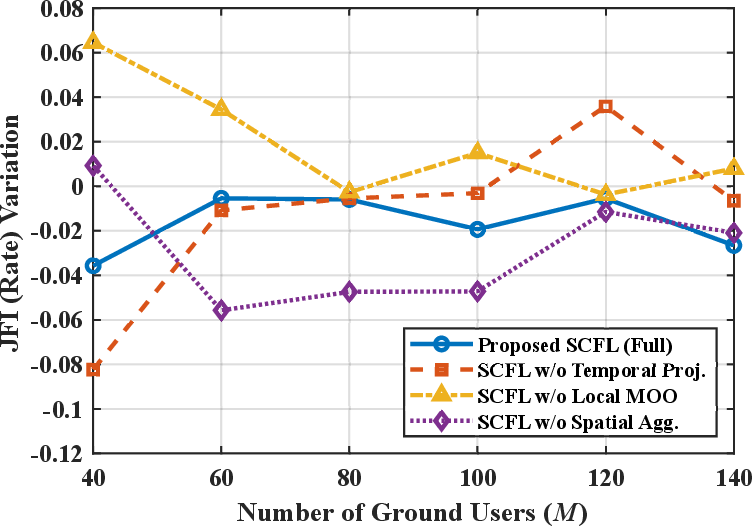}
        \caption{Variation in JFI of user data rates}
        \label{fig:ablation:JFI_rate}
    \end{subfigure}    
    
    \vspace{.2em}
    \begin{subfigure}[b]{0.5\columnwidth}
        \includegraphics[width=\linewidth]{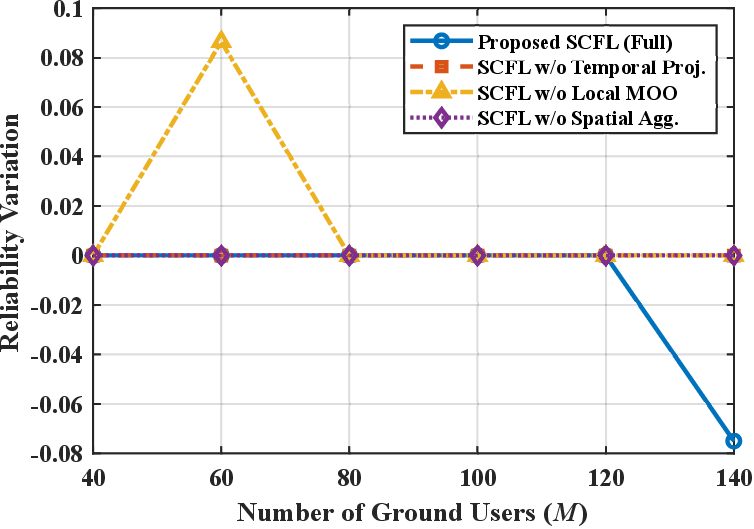}
        \caption{Variation in spatial service reliability}
        \label{fig:ablation:coverage}
    \end{subfigure}\hfill
    \begin{subfigure}[b]{0.5\columnwidth}
        \includegraphics[width=\linewidth]{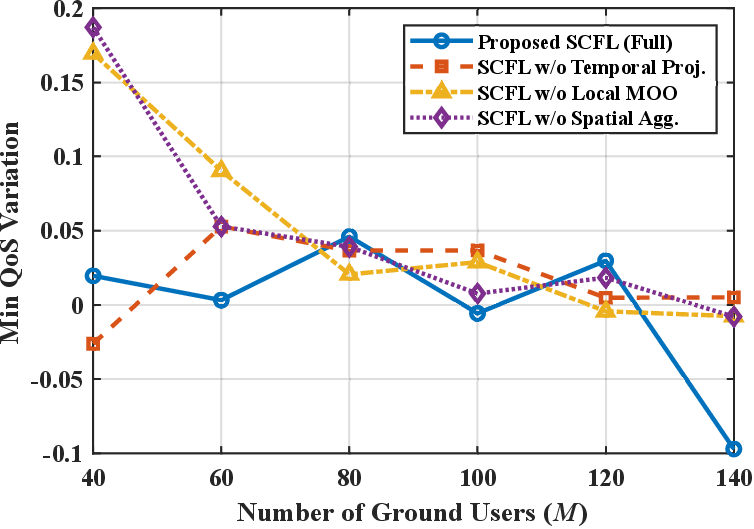}
        \caption{Variation in minimum QoS}
        \label{fig:ablation:qos}
    \end{subfigure}

    \vspace{.2em}
    \begin{subfigure}[b]{0.5\columnwidth}
        \includegraphics[width=\linewidth]{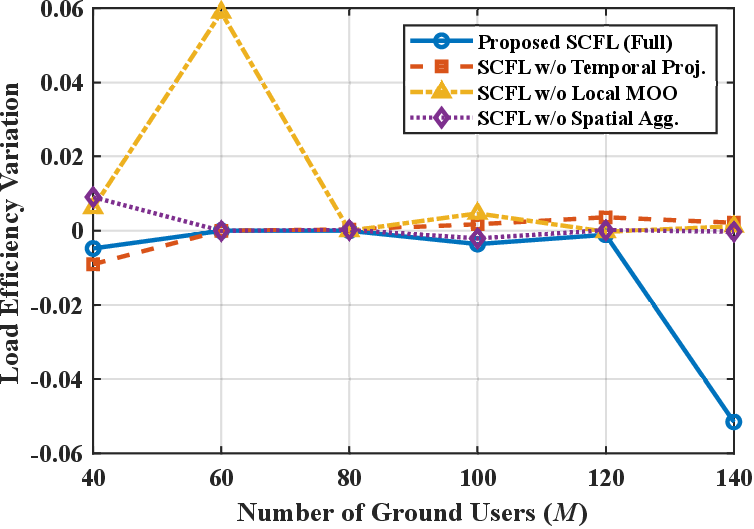}
        \caption{Variation in JFI of UAV loads}
        \label{fig:ablation:load}
    \end{subfigure}\hfill
    \begin{subfigure}[b]{0.5\columnwidth}
        \includegraphics[width=\linewidth]{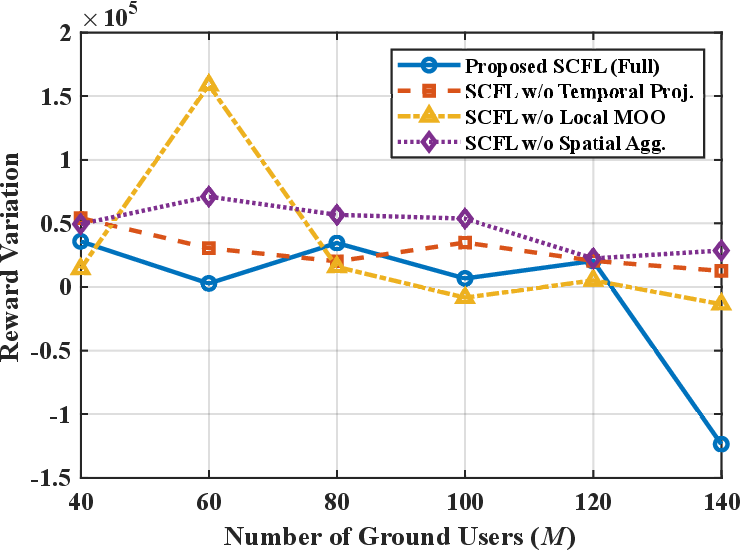}
        \caption{Variation in total system reward}
        \label{fig:ablation:reward}
    \end{subfigure}    
    \caption{Longitudinal self-degradation and ablation analysis of the proposed SCFL framework across varying user densities ($M = 40$ to $140$).}
    \label{fig:ablation}
\end{figure}

To evaluate the internal mechanisms against catastrophic forgetting, we conduct a longitudinal self-degradation and ablation analysis. Rather than comparing against external baselines with divergent initial convergence states, this evaluation compares the full SCFL framework against three ablated variants: SCFL without temporal projection (omitting global orthogonal constraints), SCFL without local MOO (bypassing local gradient decoupling), and SCFL without spatial aggregation (replacing adaptive cosine aggregation with standard federated averaging). We measure the absolute performance variation of each variant by subtracting the initial Crowded Urban baseline performance from the retained performance after sequential training. In this context, positive values ($\ge 0$) indicate retained performance at or above the initial baseline, whereas negative values ($< 0$) indicate performance regression resulting from either catastrophic forgetting or strict parameter space constraints under extreme network scales, as illustrated in Fig.~\ref{fig:ablation}.

In the moderate load regime ($M = 40$ to $120$), the full SCFL framework maintains a near-zero performance variation across all metrics. From a continual learning perspective, this near-zero variation is consistent with effective empirical retention of historical knowledge. Given the fundamentally conflicting macroscopic objectives across tasks, maintaining this baseline represents the zero-variation preservation target under the proposed first-order criterion, rather than expecting unrealistic positive backward transfer.

In contrast, the ablated variants exhibit irregular performance fluctuations. Although the variant without temporal projection occasionally achieves positive surges at specific points such as $M = 60$, these gains stem from unconstrained parameter drift rather than systematic continual adaptation. Similarly, removing local MOO causes pronounced fluctuations that disrupt the UAV load balancing in Fig.~\ref{fig:ablation:load}. Notably, as depicted in Fig.~\ref{fig:ablation:thr}, the variant without spatial aggregation exhibits apparent increases in total throughput at a substantial compromise to fairness, which is evidenced by a sharp degradation in the JFI of user data rates in Fig.~\ref{fig:ablation:JFI_rate}. This confirms that standard federated averaging is vulnerable to spatial non-IID client drift. Consequently, these observations demonstrate that SCFL operates as a conservative yet robust stability mechanism, prioritizing multi-objective equilibrium over uncontrolled parameter fluctuations.

At the extreme user density of $M = 140$, the full SCFL framework experiences performance regression, illustrating the fundamental stability-plasticity dilemma. As parameter updates become restricted, the throughput variation drops to approximately $-12.86$ Mbps in Fig.~\ref{fig:ablation:thr}, and the spatial reliability variation enters negative territory in Fig.~\ref{fig:ablation:coverage}. Similar decreases appear in the JFI of user data rates in Fig.~\ref{fig:ablation:JFI_rate} and total reward in Fig.~\ref{fig:ablation:reward}. Under moderate loads, the parameter space provides sufficient degrees of freedom to optimize new objectives within the orthogonal complement of the historical gradient direction. At $M = 140$, however, strict orthogonality limits adaptability under extreme resource contention, leading to conservative updates that prioritize historical weight protection over immediate adaptation. In contrast, the variant without temporal projection exhibits a positive performance variation (i.e., localized gain) at $M = 140$ in Fig.~\ref{fig:ablation:thr} by indiscriminately overwriting historical parameters. Although this facilitates immediate adaptation, it compromises the mathematical guarantee of knowledge retention. This trade-off establishes the empirical operating boundary induced by hard projection constraints and motivates adaptive soft-thresholding mechanisms for future work.

\section{Conclusion}
\label{sec:conclusion}

This paper addresses the stability and plasticity dilemma in non-stationary multi-UAV edge networks. We introduce the SCFL framework, which resolves conflicting gradients through a group-decoupled policy optimization mechanism and suppresses temporal interference through gradient projection. The effectiveness and robustness of the proposed framework are evaluated through comparative scalability assessments and a longitudinal self-degradation analysis. First, the horizontal comparative evaluations demonstrate that the framework achieves adaptability and synchronized fleet coordination across a sequential spatiotemporal task chain. It mitigates the reward collapse observed in off-policy baselines and provides a more favorable stability-plasticity trade-off than conventional regularization methods, sustaining a stable load distribution index of approximately 0.95 under extreme user loads. Second, the longitudinal ablation study demonstrates the role of the internal defensive mechanisms. Specifically, it confirms that adaptive spatial aggregation is essential to prevent fairness degradation caused by non-IID client drift. Under moderate network loads of up to 120 users, the framework maintains near-zero performance variation in system throughput and spatial service reliability, indicating its capability to preserve historical knowledge without computationally expensive offline retraining. Third, the evaluation under extreme congestion with 140 users provides insights into the operating boundary of hard projection constraints, where restricted parameter degrees of freedom lead to a conservative performance trade-off to protect consolidated weights. These findings establish the proposed framework as a scalable approach for autonomous and continual aerial network orchestration.






\bibliographystyle{IEEEtran}
\bibliography{IEEEabrv,reference}
\ifCLASSOPTIONcaptionsoff  \newpage \fi 


\end{document}